\pdfoutput=1
\documentclass[11pt]{article}
\usepackage{graphics,cite,amssymb,float,ifpdf}
\usepackage{amsmath, mathbbol}

\ifpdf
\usepackage[pdftex]{graphicx}
\usepackage[pdftex,dvipsnames]{xcolor}
\usepackage[final]{pdfpages}
\else
\usepackage{graphicx}
\usepackage[dvipsnames]{xcolor}
\usepackage{psfrag}
\fi
\ifpdf
\DeclareGraphicsExtensions{.pdf, .jpg, .png}
\else
\DeclareGraphicsExtensions{.eps, .jpg}
\fi

\usepackage{multirow}
\def\lsim{\;\raise0.3ex\hbox{$<$\kern-0.75em\raise-1.1ex\hbox{$\sim$}}\;}
\def\gsim{\;\raise0.3ex\hbox{$>$\kern-0.75em\raise-1.1ex\hbox{$\sim$}}\;}

\def\ben{\begin{enumerate}}  \def\een{\end{enumerate}}
\def\bit{\begin{itemize}}    \def\eit{\end{itemize}}
\def\beq{\begin{equation}}   \def\eeq{\end{equation}}
\def\ba{\begin{array}}       \def\ea{\end{array}}
\def\bea{\begin{eqnarray}}   \def\eea{\end{eqnarray}}

\begin{document}

\setcounter{footnote}{0}
\vspace*{-2.5cm}
\begin{flushright}
LPT Orsay 14-35 \\
PCCF RI 14-05

\vspace*{2mm}
\today								
\end{flushright}
\begin{center}
\vspace*{5mm}

\vspace{1cm}
{\Large\bf 
Effect of steriles states on lepton magnetic moments 
and neutrinoless double beta decay}\\
\vspace{1cm}

{\bf A. Abada$^{a}$, V. De Romeri$^{b}$ and A.M. Teixeira$^{b}$}

\vspace*{.5cm} 
$^{a}$ Laboratoire de Physique Th\'eorique, CNRS -- UMR 8627, \\
Universit\'e de Paris-Sud 11, F-91405 Orsay Cedex, France
\vspace*{.2cm} 

$^{b}$ Laboratoire de Physique Corpusculaire, CNRS/IN2P3 -- UMR 6533,\\ 
Campus des C\'ezeaux, 24 Av. des Landais, F-63177 Aubi\`ere Cedex, France

\end{center}

\vspace*{10mm}
\begin{abstract}
We address the impact of sterile fermion states on the anomalous 
magnetic moment of charged leptons, as well as
their contribution to neutrinoless 
double beta decays. We illustrate our results in a minimal, effective
extension of the Standard Model by one sterile
fermion state, and {in} a well-motivated framework of neutrino mass
generation, embedding  the Inverse Seesaw into the Standard Model.
The simple ``3+1'' effective case succeeds in alleviating the tension
related to the muon anomalous magnetic moment, albeit only at the
3$\sigma$ level, and for light sterile states ({corresponding to a }cosmologically
disfavoured regime). Interestingly, our analysis shows that a future
$0 \nu 2 \beta$ observation does not necessarily imply an inverted
hierarchy for the active neutrinos in this simple extension.
Although the Inverse Seesaw realisation here
addressed could indeed ease the tension in $(g-2)_\mu$, bounds from lepton
universality in kaon decays mostly preclude this from
happening. However, these scenarios can also have a strong impact on the
interpretation of a future $0 \nu 2 \beta$ signal regarding the hierarchy of
the active neutrino mass spectrum.

\end{abstract}
\vspace*{3mm}

\section{Introduction}
\label{sec:intro}
Extensions of the Standard Model (SM) by sterile neutrinos have received
increasing attention in recent years: in addition to their r\^ole in
numerous well motivated 
scenarios of neutrino mass generation, the existence of sterile
neutrinos is suggested by the reactor~\cite{reactor:I},
accelerator~\cite{Aguilar:2001ty,miniboone:I} 
and Gallium anomalies~\cite{gallium:I}, as well as by certain 
indications from large scale structure formation~\cite{Kusenko:2009up,Abazajian:2012ys}. 

Depending on their mass (which can range from well
below the electroweak to the Planck scale), and more importantly, 
on their mixing with the
active neutrinos, the sterile fermions can lead to non-negligible
modifications of the leptonic charged current interaction ($W \ell
\nu$), which would be manifest in electroweak (EW) interactions as a 
deviation from unitarity of the leptonic mixing
matrix~\cite{Schechter:1980gr,Gronau:1984ct}; 
in turn this will have an impact on
numerous observables, providing contributions to lepton flavour violating (LFV) 
processes~\cite{Ilakovac:1994kj,Deppisch:2004fa,Arganda:2014via}, 
leading to the violation of
lepton flavour universality (LFU)~\cite{Shrock, Nardi:1994iv,Abada:2012mc},
and enhancing rare leptonic (tau) decays,
leptonic and semi-leptonic
meson decays~\cite{Abada:2013aba}, also impacting  
invisible $Z$-boson~\cite{Akhmedov:2013hec} and Higgs boson 
decays~\cite{BhupalDev:2012zg,Das:2012ze,Cely:2012bz,Bandyopadhyay:2012px}.

Sterile neutrinos could also have a relevant r\^ole in 
flavour conserving observables, 
as is the case of electric and magnetic lepton moments, 
neutrinoless double beta decay~\cite{Deppisch:2012nb},
and several lepton number violating (LNV) decays, as for example  
$B^- \to h^+ \ell^- \ell^-$ ($h$ denoting a meson) 
among others currently being explored by the LHC
collaborations (see, for example,~\cite{LHC.LNV} and references therein). 
Although the anomalous magnetic moment of the electron, which is now
experimentally determined to an impressive precision~\cite{Hanneke:2008tm}, exhibits a
striking agreement with the SM theoretical prediction, the same does
not occur for the muon's. In fact, the
3.6$\sigma$ discrepancy~\cite{Beringer:1900zz} between the
SM prediction and the corresponding measurements, $\Delta (a_{\mu})$, 
strongly suggests that some new physics might be required in
order to reconcile theory and observation.

In this work we investigate whether the sterile fermions can provide
new contributions to the muon anomalous magnetic moment, possibly
alleviating the current tension.
For this, we consider the SM minimally extended\footnote{A
  model-independent study of an explanation for the observed
  discrepancy in the muon anomalous magnetic moment and new physics
  searches at the LHC has recently been carried in~\cite{Freitas:2014pua}.}
 by new sterile 
fields (for example right-handed neutrinos and/or other pure fermionic singlet states). 
The most minimal of these extensions consists in an effective, ad-hoc model,
where one sterile state is added to the neutral fermion
content of the SM. In this so called ``3+1" effective
framework, the additional 
state encodes the effect of a given (arbitrary) number of 
sterile fermions, and their mixings with the active (light)
neutrinos. In this approach no
assumption is made concerning the underlying mechanism of neutrino
mass generation.
On the other hand, sterile neutrinos are a crucial ingredient of many 
well-motivated mechanisms accounting for neutrino masses and mixings;
their impact on many of the
above mentioned observables is particularly important in scenarios
where the sterile states are comparatively light - as is the case 
of the $\nu$SM~\cite{Asaka:2005an}, the low-scale
type-I seesaw~\cite{Ibarra:2010xw} and the Inverse Seesaw
(ISS)~\cite{Mohapatra:1986bd}, among other possibilities. 
The ISS, where both sterile and right-handed neutrinos 
%%(with opposite lepton numbers) 
are added to the
SM field content, is particularly appealing
as it can be realised at low scales for natural values
of the neutrino Yukawa couplings. Here, and as an illustrative example
of a low-scale model of neutrino mass generation we considered a
realisation of the ISS mechanism with 3 right-handed neutrinos and 
3 additional sterile states (from now on labeled for simplicity
"ISS model").

Notice however that all these frameworks are severely constrained,
from both a theoretical and an observational point of view, and any 
realisation must comply with an extensive array of bounds. 
In addition to accommodating neutrino 
data~\cite{Tortola:2012te,Fogli:2012ua,GonzalezGarcia:2012sz,Forero:2014bxa,nufit},     
these extensions
must comply with unitarity
bounds~\cite{Antusch:2006vwa,Antusch:2008tz}, laboratory 
bounds~\cite{Lello:2012gi}, electroweak  precision
tests~\cite{delAguila:2008pw,Akhmedov:2013hec,Basso:2013jka}, 
LHC constraints (as those arising from Higgs 
decays)~\cite{BhupalDev:2012zg,Das:2012ze,Cely:2012bz,Bandyopadhyay:2012px},
bounds from rare
decays~\cite{Abada:2012mc,Abada:2013aba,Ilakovac:1994kj,Deppisch:2004fa}
as well as cosmological constraints~\cite{Smirnov:2006bu,Kusenko:2009up}.
New sources of lepton number violation can trigger neutrinoless double
beta decay (see, for example,~\cite{Deppisch:2012nb}), and 
the sterile states can contribute to the decay rate: 
the additional mixings and possible
new Majorana phases might enhance the effective mass, potentially
rendering it within experimental reach, or even leading to the
exclusion of certain regimes due to conflict with the current bounds
(the most recent results on neutrinoless double beta decay have been 
  obtained by the GERDA experiment~\cite{Agostini:2013mzu}).

Motivated by the intense experimental activity  in searching for a first signal of
neutrinoless double beta
($0 \nu 2 \beta$) decay, and in parallel to the study of the lepton
magnetic moments, we also explore in this work the impact of the sterile
fermionic states regarding neutrinoless double beta decay. 
Our analysis reveals that scenarios with sterile fermions  can indeed contribute to alleviate
the $(g-2)_\mu$ tension and, more importantly, can have a strong impact on the
interpretation of a future $0 \nu 2 \beta$ signal regarding the hierarchy of
the active neutrino mass spectrum.

Our work is organised as follows: in the subsequent section, we
address the contributions to the anomalous lepton magnetic 
moments in the presence of sterile neutrinos, also discussing their
impact for neutrinoless double beta decay. Section~\ref{sec:steriles} 
is devoted to sterile neutrino
extensions of the SM: we discuss in detail the different constraints,
and present the two minimal models which will be subsequently explored - the ``3+1''
effective model, and the ISS. Our numerical results for both 
models are collected in Section 4, where we conduct a comprehensive
analysis of the corresponding parameter spaces, and discuss the results. 
We summarise the most relevant points in the Conclusions.

\section{Sterile neutrinos and lepton magnetic moments}\label{sec:moments}

The magnetic moment of a charged lepton is given by
\begin{equation}\label{eq:magmoment:dirac}
{\overrightarrow{M}} \,= \, g_\ell \frac{e}{2m_\ell}\vec{S}\,,
\end{equation}
with $\vec{S}$, $e$ and $m_\ell$ the charged lepton's spin, electric charge and mass.
Higher order (loop) effects lead to small calculable deviations from 
the (Dirac) value $g_\ell = 2$, so that the anomalous magnetic moment
is defined as 
\begin{equation}\label{eq:magmoment:anomalous}
a_\ell\,=\,\frac{g_\ell-2}{2}\,.
\end{equation}
The experimental measurement of the  anomalous magnetic moment of the
electron differs from its SM theoretical expectation by  
\begin{equation}\label{eq:deltaae}
\Delta (a_e)\,=\, -10.5(8.1)\times 10^{-13}\,,%
\end{equation}
its theoretical prediction being dominated by QED contributions,
which have been calculated up to five loops~\cite{Aoyama:2007dv,Aoyama:2008hz}.
New physics contributions to $\Delta (a_e)$ are generally assumed to
be very small (although in~\cite{Giudice:2012ms} it was suggested  that
this observable could be used to probe and constrain new physics
scenarios, an idea also recently explored
in~\cite{Aboubrahim:2014hya}); 
in fact, the precision in the determination of $a_e$ has 
rendered it the preferred means to determine the value of the
fine-structure constant $\alpha$  (see, for example,~\cite{Mohr:2012tt}). 

On the other hand, the anomalous magnetic
moment of the muon has revealed a (yet unresolved) discrepancy between the SM 
expected value and the experimental determination.   
The current experimental  (averaged) result is given by~\cite{Mohr:2012tt}
\begin{equation}\label{eq:amu:exp}
a^\text{exp}_{\mu}\,=\,11 659 209.1(5.4)(3.3)\times 10^{-10}\,.%
\end{equation}
The SM prediction for $a^\text{SM}_{\mu}$ is generally divided into
three contributions~\cite{Passera:2004bj,Aoyama:2012wk,Czarnecki:1995sz,Gribouk:2005ee,Gnendiger:2013pva,Davier:2010nc},
\begin{equation}\label{eq:amu:SM:partial}
a^\text{SM}_{\mu}\,=\,a^\text{QED}_{\mu}+a^\text{EW}_{\mu}+a^\text{had}_{\mu}.%
\end{equation}
Hadronic (quark and gluon) loop contributions to  $a^\text{SM}_{\mu}$
are the ones most affected by theoretical uncertainties. 
By combining the different contributions, one has~\cite{Beringer:1900zz}
\begin{equation}\label{eq:amu:SM}
a^\text{SM}_{\mu}\,=\,116591803(1)(42)(26) \times 10^{-11}\,,%
\end{equation}
where the errors are due to the electroweak, lowest-order hadronic,
and higher-order hadronic contributions, respectively. 
The difference between the experimental and the theoretical values of $a_\mu$, $\Delta (a_\mu)$,  is
\begin{equation}\label{eq:deltaamu}
\Delta (a_{\mu})\,=\,a^\text{exp}_{\mu}-a^\text{SM}_{\mu}\,=\, 288 (63)(49)\times 10^{-11}\,,
\end{equation}
corresponding to a $\sim 3.6 \sigma$ deviation from the SM
prediction~\cite{Beringer:1900zz}. 

The impressive accuracy of the theoretical and experimental
results renders  $a_\mu$ a high precision observable extremely
sensitive to physics beyond the SM.
Explaining this deviation has motivated extensive studies - not only taking into
account the many possible (higher order) SM corrections, but also exploring 
new physics contributions capable of saturating the observed
discrepancy~\cite{Czarnecki:2001pv}. 
Singlet extensions of the SM have been considered to address the
$(g-2)_\mu$ discrepancy; for instance, 
standard seesaws have been investigated~\cite{Biggio:2008in}, low-scale supersymmetric seesaw models (including right-handed
neutrino superfields)~\cite{Ilakovac:2013wfa}, as well as $B-L$ 
(Inverse Seesaw) extensions of the SM~\cite{Abdallah:2011ew}, 
among many others.

Regarding the anomalous magnetic moment of the tau, 
the experimental precision~\cite{Abdallah:2003xd} is very poor compared 
with the theoretical calculation error~\cite{Eidelman:2007sb}, 
\begin{align}\label{eq:atau}
a^\text{SM}_{\tau}\,=\,117 721(5)  \times 10^{-8}\,, \nonumber \\
- 0.052 \,<\, a^\text{exp}_{\tau}\,  < \,0.013\,,
\end{align}
so that unfortunately this observable cannot in general 
be used to infer any useful information on possible new physics 
contributions.

\bigskip
\begin{figure}[h!]
\begin{center}
\includegraphics[width=50mm]{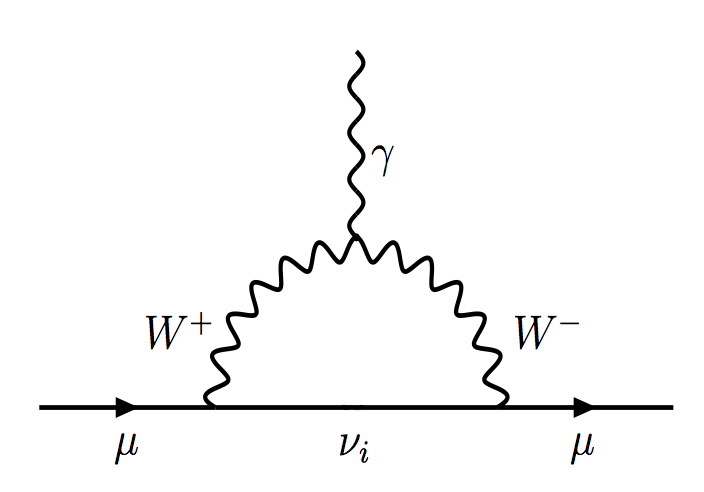}
\end{center}
\caption{One loop Feynman diagram contributing to the anomalous
  magnetic moment of charged leptons, involving the weak gauge bosons
  $W^\pm$ and the neutral fermions $\nu_i$, including the extra sterile states. } 
\label{fig:Feyndiag}
\end{figure}
In our work we address the impact of the extra sterile
neutrinos on the anomalous magnetic moment of the muon, while
considering their contribution to the $(g-2)$ of the electron, as a
potential new constraint in our analysis. 
In the SM extended by sterile fermionic states, 
the only new contribution to the anomalous magnetic moment of leptons
arises from the diagram of
Fig.~\ref{fig:Feyndiag}, where $\nu_i$ denotes the neutral fermions, including the new sterile states.
In the presence of the latter, the $W-\nu$ loop provides the following
contribution 
\begin{equation}\label{eq:amu:sterile}
a^{\nu}_{\mu}\,=\,
\frac{G_F}{\sqrt2}\,\frac{m^2_{\mu}}{8\pi^2}\,\sum^{n_\nu}_{i=1}\,U^*_{\mu
  i}U_{\mu i} \,f(x_{\nu_i}), 
\end{equation}
where $x_{\nu_i}= (m_{\nu_i}/M_W)^2$ ($n_\nu$ being the 
number of neutrino mass states, including the sterile ones)  
and $f(x)$ is given by%
\begin{equation}\label{eq:amu:sterile:fx}
f(x)= \frac{10-43x+78x^2-49x^3+4x^4+18x^3\ln (x)}{3(1-x)^4}\,.
\end{equation}
The above result is in good agreement with the corresponding one
of~\cite{Abdallah:2011ew}, derived in the framework of a $B-L$ 
extension of the SM.

\section{Sterile neutrino extensions of the SM}\label{sec:steriles}

In order to account for neutrino masses and mixings, many extensions
of the SM call upon the introduction of right-handed neutrinos (giving
rise to a Dirac mass term for the neutrinos) and/or other new
particles.  
We consider a minimal extension of the SM, an ad-hoc effective model  
(the ``3+1" effective model) as a representative case. 
There are several well-motivated neutrino mass generation
models relying on the introduction of sterile states, and in
this work we will illustrate them through 
an example of a realisation of a low scale seesaw mechanism, the {Inverse Seesaw mechanism}.
%ISS.  

\subsection{Constraints on sterile neutrino models}\label{sec:constraints}
In the framework of the SM extended by sterile fermion states, which
have a non-vanishing mixing to the active neutrinos, 
the leptonic charged currents are modified as
\begin{equation}\label{eq:cc-lag}
- \mathcal{L}_\text{cc} = \frac{g}{\sqrt{2}} U^{ji} 
\bar{\ell}_j \gamma^\mu P_L \nu_i  W_\mu^- + \, \text{c.c.}\,,
\end{equation}
where $U$ is the leptonic mixing matrix,
$i = 1, \dots, n_\nu$ denotes the physical neutrino states
and $j = 1, \dots, 3$ the flavour of the charged leptons. 
In the case of three neutrino generations,  $U$ corresponds
to  the (unitary) PMNS matrix, $U_\text{PMNS}$.
For $n_\nu >3$, the 
mixing between the left-handed leptons, which we will subsequently 
denote by $\tilde U_\text{PMNS}$, 
corresponds to a $3 \times 3$ block of $U$. 
One can parametrise the 
$\tilde U_\text{PMNS}$ mixing matrix as~\cite{FernandezMartinez:2007ms}
\begin{equation}\label{eq:U:eta:PMNS2}
U_\text{PMNS} \, \to \, \tilde U_\text{PMNS} \, = \,(\mathbb{1} - \eta)\, 
U_\text{PMNS}\,,
\end{equation}
where the  matrix $\eta$ encodes the deviation of $\tilde
U_\text{PMNS}$ from unitarity~\cite{Schechter:1980gr,Gronau:1984ct}, 
due to the presence of extra fermion states.

Many observables will be sensitive to the 
active-sterile mixings, and their current experimental values (or
bounds) will thus constrain such SM extensions. As
already mentioned, one can have LFV and LFU violating observables, 
bounds from laboratory and collider searches, among others.
Furthermore, certain
sterile mass regimes and active-sterile mixing angles are also
strongly constrained by cosmological observations. 

In what follows we proceed to discuss the most relevant constraints on
models with sterile fermions. 

\bigskip
\noindent{\bf Neutrino oscillation data:}

\noindent
The first and most important constraint to any model of massive
neutrinos is to comply with $\nu$-oscillation 
data~\cite{Tortola:2012te,Fogli:2012ua,GonzalezGarcia:2012sz,Forero:2014bxa,nufit}.
In our analysis we will consider both normal and 
inverted hierarchies for the light neutrino spectrum~\cite{Tortola:2012te}; the corresponding
best-fit intervals in the case of normal hierarchy (NH)
are\footnote{We take the local minimum in the
first octant, in agreement with~\cite{Fogli:2012ua,Forero:2014bxa}.}
\begin{align}\label{eq:neutrino.data.NH}
\sin^2 \theta_{12}\,=\, 0.32\,, 
\quad
\sin^2 \theta_{23}\,=\, 0.427 \,, 
\quad
\sin^2 \theta_{13}\,=\, 0.0246\,, \nonumber \\
\Delta m^2_{21}\,=\, 7.62\times 10^{-5} \mathrm{eV}^2\,, 
\quad
|\Delta m^2_{31}|\,=\, 2.55\times 10^{-3} \mathrm{eV}^2\, ,
\end{align}
whereas for an inverted mass hierarchy (IH) the values are
\begin{align}\label{eq:neutrino.data.IH}
\sin^2 \theta_{12}\,=\, 0.32\,, 
\quad
\sin^2 \theta_{23}\,=\, 0.6\,, 
\quad
\sin^2 \theta_{13}\,=\, 0.025\,, \nonumber \\
\Delta m^2_{21}\,=\, 7.62\times 10^{-5} \mathrm{eV}^2\,, 
\quad
|\Delta m^2_{31}|\,=\, 2.43\times 10^{-3} \mathrm{eV}^2\, . 
\end{align}
The value of the CP violating Dirac phase $\delta$ is still
undetermined, although the complementarity of accelerator and reactor
neutrino data starts reflecting in a better sensitivity to 
$\delta$~\cite{Capozzi:2013csa,Forero:2014bxa} 
(and to the hierarchy of the light neutrino spectrum).

It is worth noticing that the CP violating phases of the $U_\text{PMNS}$, 
as well as the possible new phases that will be
present in extensions of the SM by sterile neutrinos, will
also contribute to the electric dipole moments (EDM) of leptons; however, in
these minimal extensions, lepton EDMs would only receive contributions
at the 2-loop level. We do not address these CP violating (CPV) 
observables in the present analysis.

\bigskip
\noindent{\bf Unitarity constraints:}

\noindent
The introduction of fermionic sterile states can give
rise to non-standard neutrino interactions with matter.  
Bounds on the non-unitarity of the matrix $\eta$ (cf. Eq.~(\ref{eq:U:eta:PMNS2})),
have been derived in~\cite{Antusch:2008tz}  
by means of an effective theory approach.  
We apply these bounds in our numerical analysis in the regimes for which the
latter approach is valid, generically for sterile masses above the GeV, but 
below the EW scale (at least in extensions calling for more than one
sterile state).

\bigskip
\noindent{\bf Electroweak precision data:}

\noindent
The addition of (fermion) singlets to the SM with a sizeable active-sterile
mixing can affect electroweak precision observables at
tree-level (charged currents) and at higher order as well. 
In particular, the  non-unitarity of the active neutrino
mixing matrix, Eq.~(\ref{eq:U:eta:PMNS2}), implies that the 
couplings of the active neutrinos to the $Z$ and $W$
bosons are suppressed with respect to their SM values. 
In the presence of singlet neutrinos, electroweak precision constraints were
first addressed in~\cite{delAguila:2008pw} with an effective
approach (therefore valid only for multi-TeV singlet states). 
More recently, the effects of the sterile neutrinos on 
the invisible $Z$-decay width have been 
discussed in~\cite{Akhmedov:2013hec,Basso:2013jka,Abada:2013aba}, and it
has been shown that $\Gamma(Z \to \nu \nu)$ can be reduced 
with respect to the SM prediction. Complying with  LEP results
on $\Gamma(Z \to \nu \nu)$~\cite{Beringer:1900zz} also constrains
these sterile neutrino extensions.

\bigskip
\noindent{\bf LHC constraints:}

\noindent
The new interactions in the leptonic sector can also alter the
Higgs boson phenomenology: the presence of a new decay channel for the
Higgs boson, with heavy 
%%Majorana 
neutrinos in the final state, 
can enlarge the total Higgs decay width, 
thus lowering the predicted SM decay branching ratios. 
Therefore, constraints on sterile neutrinos are also 
derived from Higgs decays.  LHC data already allows to constrain
regimes where the sterile states are below the Higgs mass, 
due to the potential Higgs decays to an active and a heavier, 
mostly sterile, neutrino. In our analysis we apply the constraints derived 
in~\cite{BhupalDev:2012zg,Cely:2012bz,Bandyopadhyay:2012px}.

\bigskip
\noindent{\bf Leptonic and semileptonic meson decays:}

\noindent
Further constraints arise from decays of pseudoscalar mesons
$K$, $D$, $D_s$ and $B$, with one or two neutrinos in
the final state (see~\cite{Goudzovski:2011tc,Lazzeroni:2012cx} for
kaon decays,~\cite{Naik:2009tk,Li:2011nij} for $D$ and $D_s$ decay
rates, and~\cite{Aubert:2007xj,Adachi:2012mm} for $B$-meson observations).
In the framework of the SM extended by sterile neutrinos, 
these decays have recently been addressed in~\cite{Abada:2012mc,Abada:2013aba}. 
The dominant contributions to these processes arise from tree-level
$W$ mediated exchanges (a consequence of the 
modified vertex $W\ell \nu$ due to the presence of the sterile states). 
As will be discussed in the following section, among the distinct 
constraints derived from meson decays, the most
severe bounds are due to the violation of lepton universality in leptonic kaon
decays, parametrised by 
$\Delta r_K$, 
\begin{equation}\label{eq:deltarK}
\Delta r_K \, \equiv \, \frac{R_K^\text{exp}}{R_K^\text{SM}} - 1\, 
\quad \text{where} \quad
R_K \, \equiv \,\frac{\Gamma (K^+ \to e^+ \nu)}{\Gamma (K^+ \to \mu^+
  \nu)}\,,
\end{equation}
and whose current value (comparison of theoretical 
analyses~\cite{Cirigliano:2007xi,Finkemeier:1995gi} with
the recent measurements from the NA62 
collaboration~\cite{Goudzovski:2011tc}) is 
\begin{equation}\label{eq:deltarK:value} 
\Delta r_K \, = \, (4 \pm 4 )\, \times\, 10^{-3}\,.
\end{equation}
This observable can receive significant contributions from the sterile 
states, due to the new phase space factors and as a result of
deviations from unitarity, when the sterile mixings to the 
active neutrinos are sizeable~\cite{Shrock,Abada:2012mc,Abada:2013aba}. 

\bigskip
\noindent{\bf Laboratory searches:}

\noindent
Robust laboratory bounds on the sterile neutrino masses and their
mixings with the active ones  
can be inferred from negative searches for monochromatic lines in the
spectrum of muons from  $\pi^\pm \to \mu^\pm \nu$
decays~\cite{Kusenko:2009up,Atre:2009rg}. 
The absence of such a signal imposes 
stringent limits for sterile neutrinos with masses in the MeV-GeV range.

\bigskip
\noindent{\bf Lepton flavour violation:}

\noindent
Non-negligible active-sterile mixings will affect 
charged lepton violating (cLFV) processes, leading to rates
potentially larger than current bounds, through the enlarged leptonic
mixing matrix. 
The most stringent bound on sterile neutrinos from cLFV 
processes~\cite{Ilakovac:1994kj,Deppisch:2004fa} comes from the
search for the radiative $\mu \to e \gamma$ decay~\cite{Adam:2013mnn}.

\bigskip
\noindent{\bf Neutrinoless double beta decay:}

\noindent
The introduction of singlet neutrinos with Majorana masses allows for 
new processes like LNV interactions.
Among these, neutrinoless double beta decay remains the most important
one. This process is being actively searched for by several
experiments, by means of the  best performing detector techniques: 
among others, GERDA~\cite{Agostini:2013mzu}, 
EXO-200~\cite{Auger:2012ar,Albert:2014awa} and
KamLAND-ZEN~\cite{Gando:2012zm}, have all set strong bounds on the 
effective mass, $| m_{ee}| $, to which the amplitude of $0\nu 2 \beta$  
process is proportional. 
The sensitivities of the current experiments put a limit on the effective
neutrino Majorana mass in the range 
\beq
| m_{ee}| \lsim 140\text { meV} - 700\text { meV}\,.
\eeq
In Table~\ref{tab:nulesssensitivities}, we summarise the future 
sensitivity of ongoing and planned $0\nu 2 \beta$ experiments.

\begin{table}[h!]
\begin{center}
{\begin{tabular}{| l | l | c |}  \hline                       
Experiment & Ref. &  $ |m_{ee} | $ (eV) \\
  \hline                       
 EXO-200 (4 yr) & \cite{Auger:2012ar,Albert:2014awa} & 0.075 - 0.2  \\
nEXO (5 yr)  & \cite{DeliaTosionbehalfoftheEXO:2014zza}& 0.012 - 0.029  \\
 nEXO (5 yr + 5 yr w/ Ba tagging) &
 \cite{DeliaTosionbehalfoftheEXO:2014zza} & 0.005 - 0.011  \\ 
KamLAND-Zen (300~kg, 3 yr)& \cite{Gando:2012zm}  & 0.045 - 0.11 \\
GERDA  phase II & \cite{Agostini:2013mzu} & 0.09 - 0.29 \\
CUORE (5 yr) & \cite{Gorla:2012gd,Artusa:2014lgv} & 0.051 - 0.133 \\
SNO+  & \cite{Hartnell:2012qd} & 0.07 - 0.14 \\
SuperNEMO & \cite{Barabash:2012gc} & 0.05 - 0.15 \\
NEXT & \cite{Granena:2009it,Gomez-Cadenas:2013lta}& 0.03 - 0.1 \\
MAJORANA demo. & \cite{Wilkerson:2012ga} & 0.06 - 0.17 \\
  \hline                       
\end{tabular}
}
\caption{Future sensitivity of several $0\nu 2 \beta$ experiments.}
\label{tab:nulesssensitivities}
\end{center}
\end{table}

\noindent
This observable will be addressed in detail when we discuss each of
the sterile neutrino models explored in our study.

\bigskip
\noindent{\bf Cosmological bounds:}

\noindent
Sterile neutrinos with a mass below the TeV are subject to strong
constraints from a number of
cosmological observations~\cite{Smirnov:2006bu,Kusenko:2009up}. The
sterile states play an important r\^ole in cosmology and
astrophysics, in particular in Big Bang Nucleosynthesis and  Large Scale
Structure formation. Moreover, a sterile neutrino with a mass $\sim$~keV  
may be a viable dark matter candidate, for instance 
offering a possible explanation  
to the observed X-ray line in cluster galaxy spectra at $\sim
3.5$~keV~\cite{Bulbul:2014sua,Boyarsky:2014jta}, to  
the origin of pulsar kicks, or even to the baryon asymmetry of the
Universe (for a review see~\cite{Abazajian:2012ys}). 

These cosmological limits are in general derived by
assuming the minimal possible abundance (in agreement with neutrino
oscillations) of sterile neutrinos in halos consistent with standard
cosmology. 
Nevertheless, as argued in~\cite{Gelmini:2008fq}, the possibility of a
non-standard cosmology with a very low reheating temperature, or a
scenario where the sterile 
neutrinos couple to  a dark sector~\cite{Dasgupta:2013zpn}, could
allow to evade some of the above bounds. For this reason,  and aiming at being
conservative, in our numerical study we will allow for the violation
of these cosmological bounds in some scenarios, explicitly stating it.

\subsection{Effective SM extension: ``3+1'' model}\label{sec:EFF}

A first approach to address the impact of sterile fermions on the
magnetic moments of leptons is to consider a minimal
\textit{effective} model with three light active neutrinos and one extra
sterile Majorana neutrino. In this approach no
assumption is made concerning the underlying mechanism responsible for 
the origin of neutrino masses and mixings.
The extension of the SM by the extra state introduces additional
degrees of freedom: its mass $m_4$, three new (active-sterile) mixing angles
$\theta_{i4}$, two new (Dirac) CP violating phases and one extra
Majorana phase.

Concerning the anomalous lepton magnetic moments, the sum in
Eq.~(\ref{eq:amu:sterile}) extends to $n_\nu=4$, while 
the effective neutrino mass $m_{ee}$, determining the 
amplitude of the neutrinoless double beta decay rate, is given
by~\cite{Blennow:2010th}: 
\begin{equation} \label{eq:22bbdecay}
 m_{ee}\,\simeq \,\sum_{i=1}^4 U_{ei}^2 \,p^2
\frac{m_i}{p^2-m_i^2} \simeq 
\left(\sum_{i=1}^3 U_{ei}^2\, m_{\nu_i}\right)\, 
+ p^2 \, U_{e4}^2 \,\frac{m_4}{p^2-m_4^2}\,,
\end{equation}
where $p^2 \simeq - (100 \mbox{ MeV})^2$ is the virtual momentum of
the neutrino (an average estimate over different values depending on
the decaying nucleus).

\subsection{The Inverse Seesaw scenario}\label{sec:ISS}
In order to investigate the impact of the sterile neutrinos on the
lepton magnetic moments in the concrete framework of a neutrino mass
generation mechanism, we have considered the Inverse Seesaw 
mechanism~\cite{Mohapatra:1986bd}. The ISS scenario 
is an appealing extension of the SM, which allows to accommodate
neutrino data with natural values of the Yukawa couplings for a
comparatively low seesaw scale. In turn, this offers the possibility
of having sizeable mixings between the active
neutrinos and the additional sterile states, thus rendering the model
phenomenologically rich. 

The ISS requires the introduction of $n_R \ge
2$ generations\footnote{This lower value is required in order to account for the
  active neutrino masses and mixings. The most minimal Inverse Seesaw 
  realisation~\cite{Abada:2014vea} consists in the addition of two
  right-handed and two sterile neutrinos to the SM content.}
of right-handed (RH) neutrinos $\nu_R$ and $n_X$
generations of extra $SU(2)$ singlets fermions $X$ (such that 
$n_R+n_X = N_s$), both with lepton number $L=+1$~\cite{Mohapatra:1986bd}.
In our analysis we will consider a realisation with $n_R = n_X = 3$.

In the ISS, the SM Lagrangian is extended as
\begin{equation}
\mathcal{L}_\text{ISS} \,=\, 
\mathcal{L}_\text{SM} - Y^{\nu}_{ij}\, \bar{\nu}_{R i} \,\tilde{H}^\dagger  \,L_j 
- {M_R}_{ij} \, \bar{\nu}_{R i}\, X_j - 
\frac{1}{2} {\mu_X}_{ij} \,\bar{X}^c_i \,X_j + \, \text{h.c.}\,,
\end{equation}
where $i,j = 1,2,3$ are generation indices and $\tilde{H} = i \sigma_2
H^*$.
After EW symmetry breaking, the (symmetric) $9\times9$ neutrino mass matrix 
$\mathcal{M}$ is given in the $(\nu_L,{\nu^c_R},X)^T$ basis by
\begin{eqnarray}
{\cal M}&=&\left(
\begin{array}{ccc}
0 & m^{T}_D & 0 \\
m_D & 0 & M_R \\
0 & M^{T}_R & \mu_X \\
\end{array}\right) \, .
\label{eq:ISS:M9}
\end{eqnarray}
Notice that $U(1)_L$ (i.e., lepton number) is broken
only by the non-zero Majorana mass term $\mu_{X}$, while the Dirac-type 
right-handed neutrino mass term $M_{R}$ conserves lepton number. In
the above,  
$m_D= \frac{1}{\sqrt 2} Y^\nu v$ is the Dirac mass term,  
$v$ being the vacuum expectation value of the SM Higgs boson.  
Assuming $\mu_X \ll m_D \ll M_R$, the
diagonalization of ${\cal M}$ leads to an effective Majorana mass
matrix for the active (light) neutrinos~\cite{GonzalezGarcia:1988rw},
\begin{equation}\label{eq:nu}
m_\nu \,\simeq \,{m_D^T\, M_R^{T}}^{-1} \,\mu_X \,M_R^{-1}\, m_D \, .
\end{equation}
The remaining (mostly) sterile states form nearly degenerate pseudo-Dirac
pairs, with masses 
\begin{equation}\label{eq:ISS:pseudodirac}
 m_{S_\pm}=\pm \sqrt{M_{R}^2+m_{D}^2} + \frac{M_{R}^2 \,\mu_X}{2\,
   (m_{D}^2+M_{R}^2)}\,.  
\end{equation}
For the purpose of our analysis it is useful to define 
$M = M_R\, \mu_X^{-1} \,M_R^T$, which is diagonalized by the matrix 
$D$ as $D M D^T = \hat{M}$. The eigenvalues of $M$ are thus the
entries of the diagonal matrix $\hat M$. It is also convenient to
generalize the Casas-Ibarra parametrisation~\cite{Casas:2001sr}, 
which allows to write the neutrino Yukawa couplings $Y^\nu$ as
\begin{equation}\label{eq:YvcasasI}
Y^\nu \,= \,\frac{\sqrt{2}}{v} \, D^\dagger \, \sqrt{\hat M} \, R \,
\sqrt{{\hat m}_\nu} \, U_\text{PMNS}^\dagger \, ,
\end{equation}
where $\sqrt{{\hat m}_\nu}$ is a diagonal matrix containing the square
roots of the three light neutrino mass eigenvalues $m_\nu$. $R$ is an
arbitrary $3 \times 3$ complex 
orthogonal matrix, parametrized by $3$ complex angles, which encodes the
remaining degrees of freedom. (Without loss of generality, we can work
in the basis where $M_R$ is a real diagonal matrix, as are the
charged lepton Yukawa couplings.)
The neutrino mass matrix is then diagonalized by a $9\times 9$
unitary mixing matrix $U$ as $U^T \mathcal{M} U 
= \text{diag}(m_i)$. In the basis where the charged lepton mass matrix is
diagonal, the leptonic mixing matrix is given by the
rectangular $3 \times 9$ sub-matrix corresponding to the first three
columns of $U$, with the $3 \times 3$ block corresponding to the 
(non-unitary)\footnote{We refer 
to~\cite{Forero:2011pc,Malinsky:2009gw,Dev:2009aw} for earlier studies on
non-unitarity effects in the Inverse Seesaw.} $\tilde U_\text{PMNS}$.

\bigskip
While in the ``3+1'' effective model, four states  (three active and one
sterile)  contributed to the sum of 
Eq.~(\ref{eq:amu:sterile}), in the ISS the sum includes 9 mass
eigenstates, and reflects the possibility of having the 
additional (mostly) sterile states present in the loop of
Fig.~\ref{fig:Feyndiag}, thus significantly contributing to the
anomalous lepton moments.

Concerning $0 \nu 2 \beta$ decays, the new sterile states will also contribute:
their spectrum corresponding to three pseudo-Dirac pairs, 
cf. Eq.~(\ref{eq:ISS:pseudodirac}), 
the effective neutrino mass $m_{ee}$ 
is now given by
\begin{eqnarray} \label{eq:22bbdecay:ISS}
m_{ee}&\,\simeq\,&\sum_{i=1}^9 U_{ei}^2 \,p^2
\frac{m_i}{p^2-m_i^2}\,\simeq \,\left(\sum_{i=1}^3 \,U_{ei}^2\,
m_{\nu_i}\right) + \nonumber\\  
&&+\, p^2 \,\Big(- U_{e4}^2 \,\frac{|m_4|}{p^2-m_4^2}+U_{e5}^2\,
\frac{|m_5|}{p^2-m_5^2}-U_{e6}^2 \,\frac{|m_6|}{p^2-m_6^2}+
\nonumber\\ 
&& +\, U_{e7}^2 \,\frac{|m_7|}{p^2-m_7^2}-U_{e8}^2
\,\frac{|m_8|}{p^2-m_8^2}+U_{e9}^2 \,\frac{|m_9|}{p^2-m_9^2}\,\Big)\,, 
\end{eqnarray}
with, as before, $p^2 \simeq - (100 \mbox{ MeV})^2$ the virtual momentum of the
neutrino.

\section{Analysis and discussion}
The aim of this study is to evaluate the impact of the new
sterile states on the leptonic anomalous magnetic moment, as well as
on the neutrinoless double beta decay effective mass, for the two models
previously discussed - the ``3+1'' effective model, and the ISS with 3
right-handed neutrinos and 3 extra sterile states. 
As mentioned in Section~\ref{sec:constraints},
we will apply in each case all relevant constraints.
In the different plots illustrating our study, and 
for the sake of completeness, we will nonetheless show all the
solutions, denoting by different shades of grey the points not 
complying with the various bounds. 
An exception concerns the cosmological constraints:  
as noticed for example in~\cite{Gelmini:2008fq}, the cosmological bounds
could be modified and eventually evaded by considering a
non-standard cosmology. Therefore, and throughout the subsequent
discussion, we will not discard solutions in disagreement with the 
cosmological bounds of~\cite{Kusenko:2009up}, highlighting them {with} a
distinctive colour scheme, namely with red coloured points. 

As a convenient means to illustrate the effect of the new
active-sterile mixings (corresponding to a deviation from unitarity of
the $\tilde U_\text{PMNS}$) we have introduced the invariant quantity
$\tilde \eta$, defined as
\begin{equation}\label{eq:def:etatilde}
\tilde \eta = 1 - |\text{Det}(\tilde U_\text{PMNS})| \, .
\end{equation}
In our analysis we address 
both cases of normal and inverted hierarchies for the light
neutrino spectrum, and we fully explore the parameter space 
for each of the models considered,
including the new (Dirac and Majorana) CP violating phases. As already
stated, we do not address the (2-loop) contributions to the leptonic
EDMs.

\subsection{Results: ``3+1'' effective model} \label{subsec:EFFres}

For both NH and IH light neutrino spectra, we 
scan over the sterile neutrino mass in the range
\begin{equation}\label{eq:eff:scan:1}
3 \times 10^{-11} \text{ GeV} \lesssim m_4 \lesssim 10^3 \text{ GeV}\ ,
\end{equation}
and over the active-sterile mixing angles $\theta_{i4}$ ($i=1,2,3$),
which are randomly taken between  0 and 2$\pi$. We also consider the
effect of all CPV phases, which are likewise randomly varied in $[0, 2\pi]$.

\bigskip
\noindent
{\bf Anomalous magnetic moment of the electron}

\begin{figure}[!h]
\begin{center}
\includegraphics[width=60mm, angle=270]{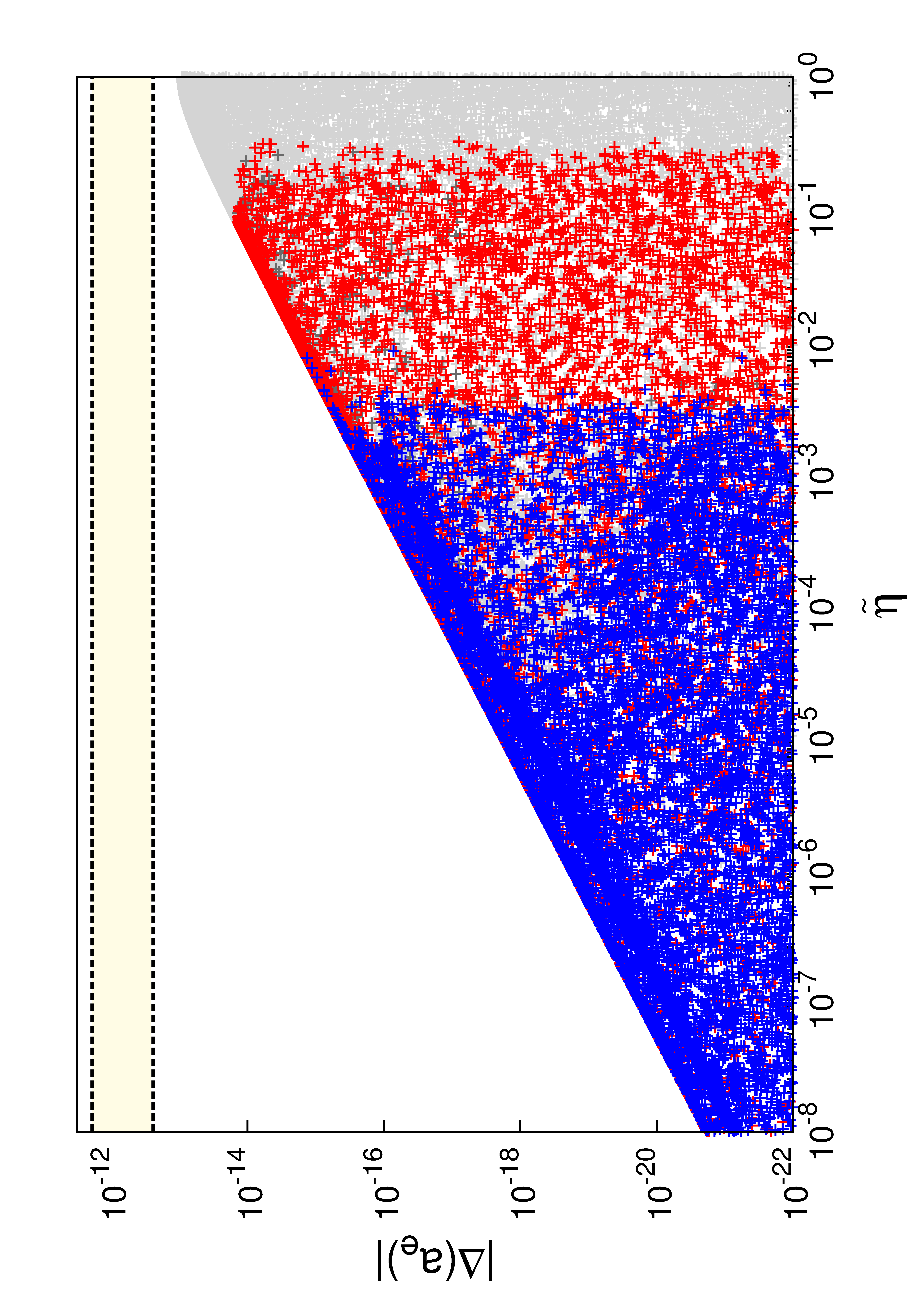}
\end{center}
\caption{Effective case: anomalous magnetic moment of the electron as
  a function of $\tilde \eta$ (see Eq.~(\ref{eq:def:etatilde})), 
  for a NH light neutrino spectrum. Blue
  points are in agreement with cosmological bounds, while the red ones
  would require considering a non-standard cosmology. In grey we
  denote points already excluded by other (non-cosmological) bounds
  (see text for a description of the grey-shade scheme). The beige coloured
  band denotes the 1$\sigma$ interval, cf. Eq.~(\ref{eq:deltaae}). 
  The underlying scan is described in the text. }\label{fig:3+1:deltaae:nh}
\end{figure}

\noindent
We first consider the potential impact of $\Delta (a_e)$
as an additional constraint
on  the parameter space of the ``3+1'' effective extension of the SM.  
As can be seen on Fig.~\ref{fig:3+1:deltaae:nh}, which illustrates
this observable for the case of a normal hierarchy in the light
neutrino spectrum, the new contributions to $|\Delta (a_e)|$ do augment
with increasing deviations from unitarity of the $\tilde
U_\text{PMNS}$ matrix. However, even for the largest values of
$\tilde \eta$ - which are excluded due to the violation of several
bounds (the most important one being neutrino oscillation data) 
- the ``3+1'' effective model remains
short of the 1$\sigma$ bound. As already mentioned in
Section~\ref{sec:constraints}, we display in red the points that 
are typically disfavoured from standard cosmology arguments
(notice that the large $\tilde \eta$  regime in general corresponds to
this case). The grey-shades correspond to at least: 
failure to comply with $\nu$-oscillation data,
bounds from EW precision data or LHC bounds
(light grey);  violation of laboratory bounds or constraints from
rare leptonic meson decays (grey); confict with bounds from 
radiative $\mu \to e \gamma$ decays, 
neutrinoless double beta decays or invisible $Z$-boson width
(dark grey). 

\bigskip
\noindent
{\bf Anomalous magnetic moment of the muon}

\begin{figure}[!h]
\begin{center}
\begin{tabular}{cc}
\includegraphics[width=55mm,angle= 270]{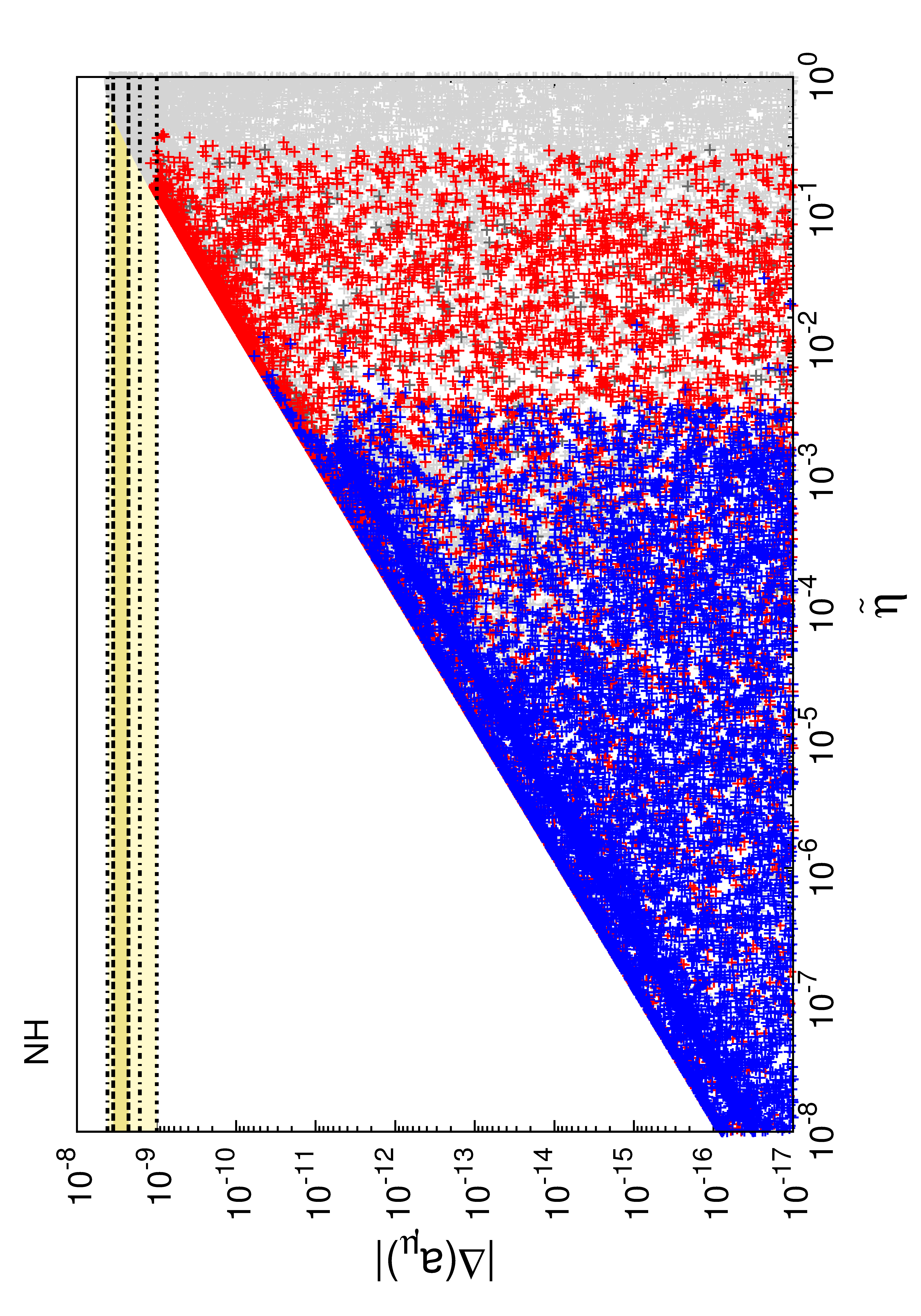}
&
\includegraphics[width=55mm,angle= 270]{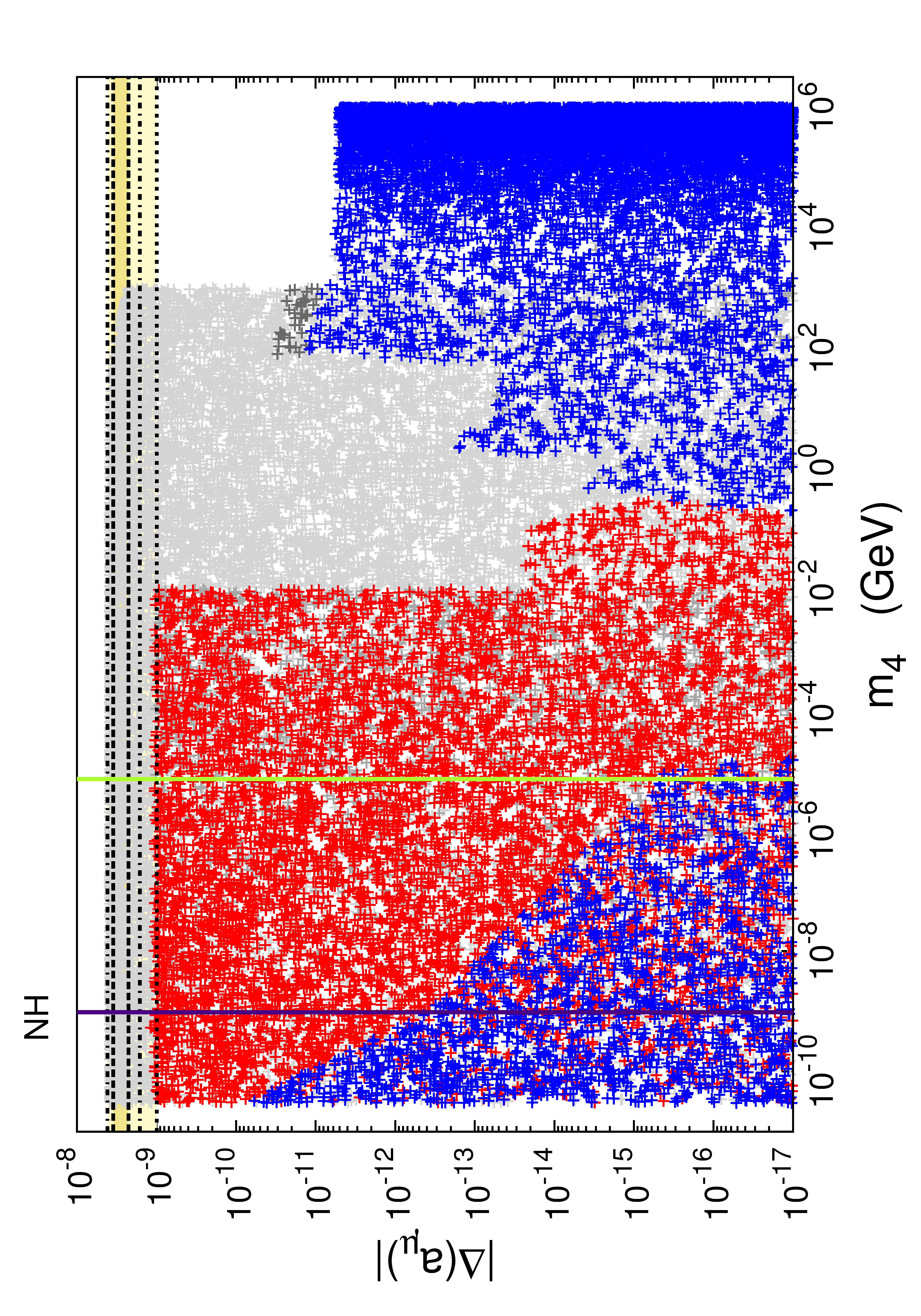}
\end{tabular}
\end{center}
\caption{Effective case: anomalous magnetic moment of the muon as
  a function of $\tilde \eta$ (left) and of the mass of the sterile
  neutrino $m_4$ (right), for a NH light neutrino spectrum. The
  horizontal dashed lines delimit the surfaces corresponding to the 
  1$\sigma$, 2$\sigma$ and 3$\sigma$ intervals for $|\Delta (a_\mu)|$
  (darker to lighter shades of beige), see Eq.~(\ref{eq:deltaamu}). 
  The vertical violet (green)
  line corresponds to the best fit value of $m_4$ which would account
  for the reactor anomaly (to the 3.5~keV line from decaying warm dark matter). 
  Otherwise, colour scheme as in
  Fig.~\ref{fig:3+1:deltaae:nh}.}\label{fig:3+1:deltaamu:nh:eta:m4}
\end{figure}

\noindent
We proceed to investigate whether this minimal extension of the SM
with sterile neutrinos can contribute to alleviate the tension between the 
experimental measurement  and the SM prediction for $(g-2)_\mu$, as discussed in
Section~\ref{sec:moments}. 
On Fig.~\ref{fig:3+1:deltaamu:nh:eta:m4}, we display the contribution
of the ``3+1'' effective model, for the case of a normal hierarchical
light neutrino spectrum. Although we do not explicitly display them
here, phenomenological equivalent results have been obtained for the
case of an inverted hierarchy in the active neutrino
spectrum. 

The left panel of Fig.~\ref{fig:3+1:deltaamu:nh:eta:m4} reveals that this simple
effective extension can account for a contribution capable of slightly 
alleviating the tension between theory and experiment, albeit close to
the viability limit of the model (i.e., large values of $\tilde
\eta$), and only for points disfavoured by cosmological observations. 
Moreover, these contributions are associated to light sterile
states: as can be seen from the right panel of
Fig.~\ref{fig:3+1:deltaamu:nh:eta:m4}, contributions within the
3$\sigma$ interval correspond to $m_4 \lesssim 10^{-2}$ GeV. The two
vertical lines of the right panel denote the values of the mostly (lightest) 
sterile state mass that would allow to address issues strongly
motivating these extensions of the SM: (i) $m_4 \simeq 1$ eV, the best
fit value for a sterile state accounting for the reactor
anomaly~\cite{Kopp:2013vaa}; (ii) $m_4 \sim 7$ keV, as required to explain the 
3.5~keV line in the X-ray spectra of galaxy clusters from the decay of 
a warm dark matter candidate~\cite{Bulbul:2014sua,Boyarsky:2014jta}. 

We stress that this is a simple, minimal extension and that the
analysis relies on a first order (one-loop) computation of the
observable. A more complete computation, including higher order
(leptonic and hadronic) contributions could further reduce the
present discrepancy.

\bigskip
\noindent
{\bf Anomalous magnetic moment of the tau}

\noindent
We have also considered the contribution of the ``3+1" effective model to
the anomalous magnetic moment of the tau lepton. Throughout the
investigated parameter space, we found that the new contributions
saturate at 
\begin{equation}\label{eq:res:3+1:deltaatau}
|\Delta (a_{\tau})| \lesssim 10^{-5}\,,
\end{equation}
and are thus clearly beyond experimental reach.

\bigskip
\noindent
{\bf Neutrinoless double beta decay}

\begin{figure}[!h]
\begin{center}
\begin{tabular}{cc}
\includegraphics[width=55mm,angle= 270]{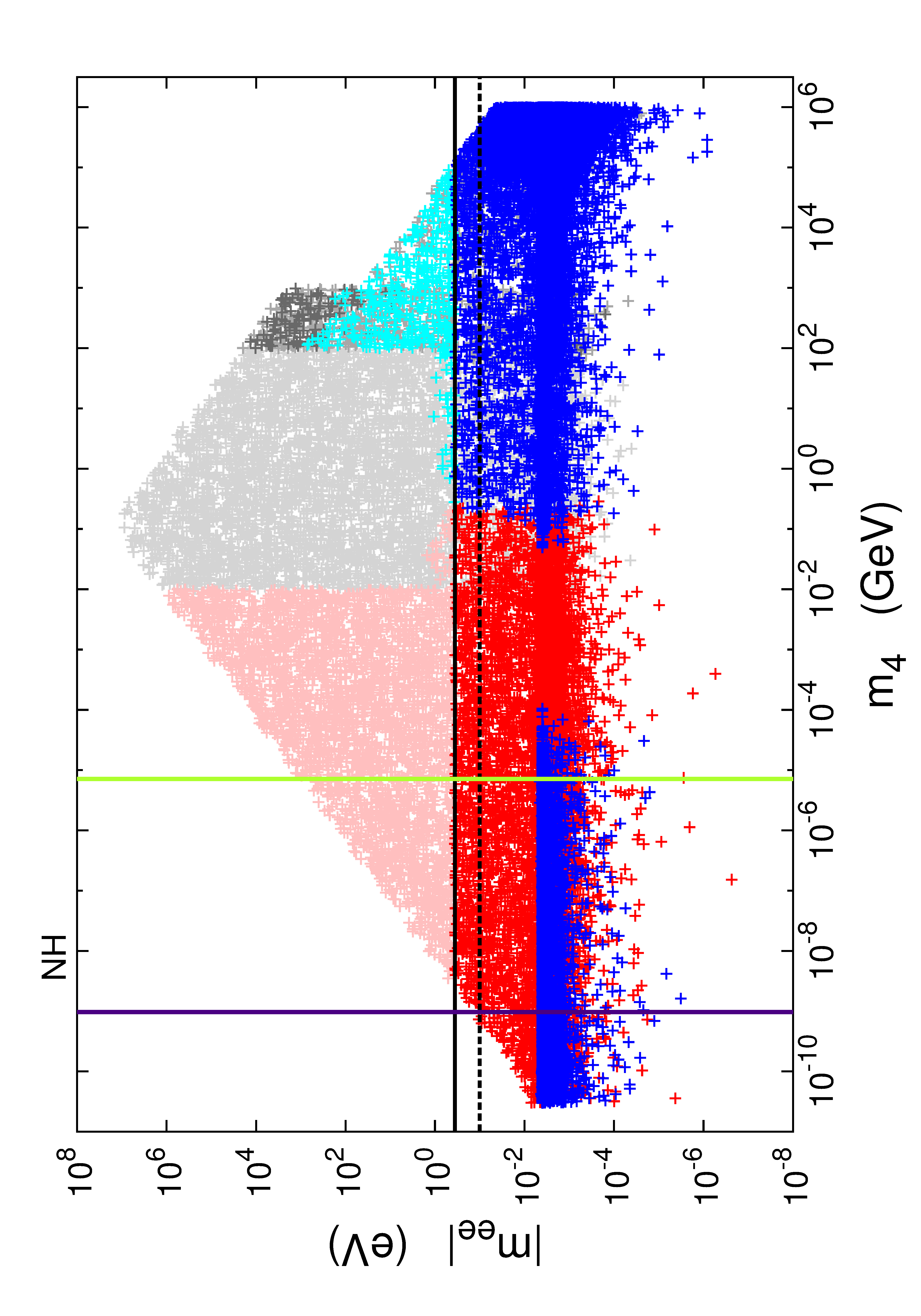}
&
\includegraphics[width=55mm,angle= 270]{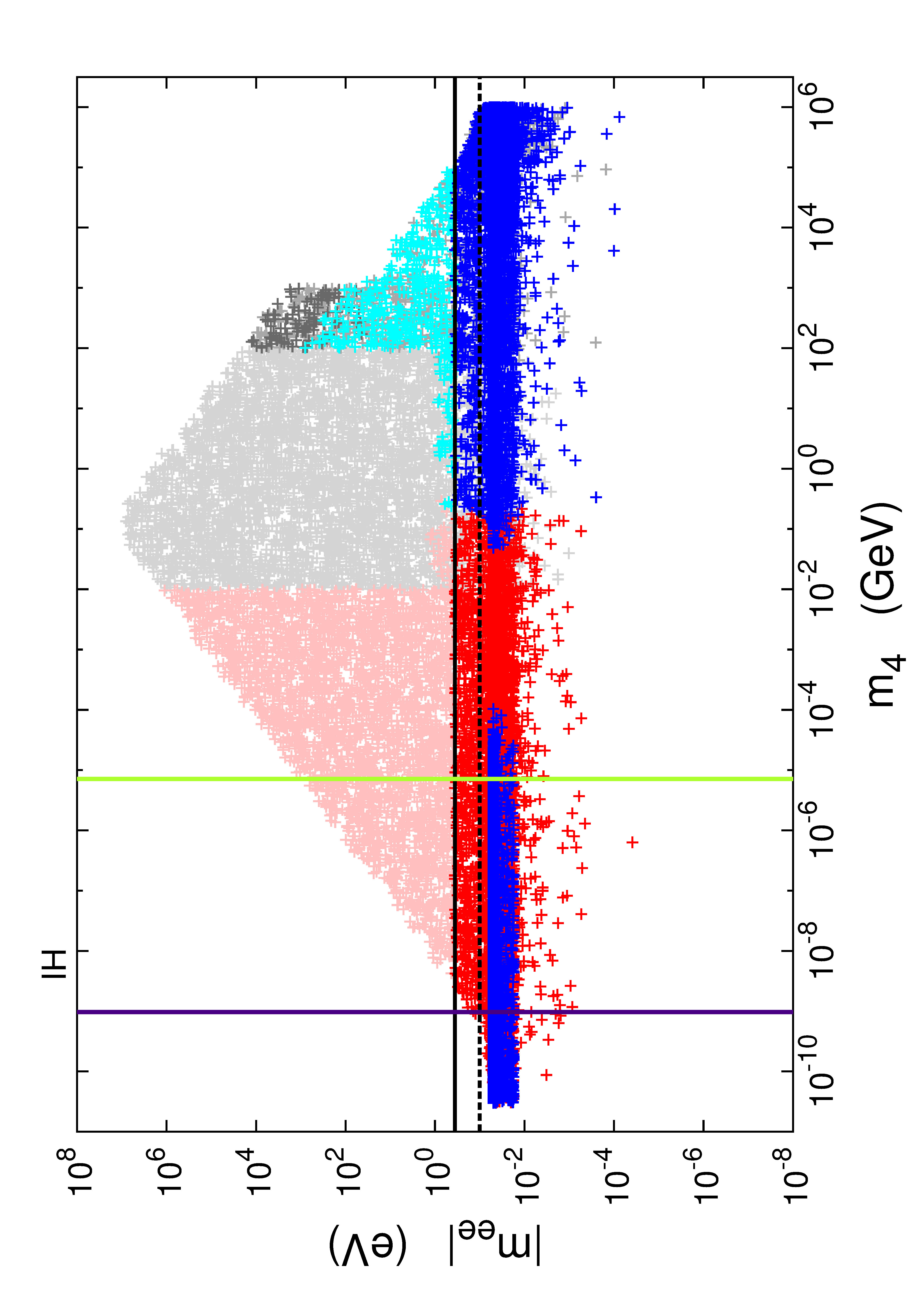}
\end{tabular}
\end{center}
\caption{Effective case: effective mass, $|m_{ee}|$ (in eV), as a
  function of the mass of the sterile
  neutrino $m_4$, for NH (left) and IH (right) light neutrino
  spectra. The solid (dashed) horizontal line denotes the current
  bound (expected future sensitivity), see
  Table~\ref{tab:nulesssensitivities}, while the vertical lines 
  correspond to the values of $m_4$ which would account
  for the reactor anomaly and explain the 
  3.5 keV line from decaying warm dark matter. Blue
  points are in agreement with cosmological bounds, while the red ones
  would require considering a non-standard cosmology. Cyan/pink
  respectively correspond to cosmologically favoured/disfavoured
  points which are excluded due to conflict with bounds from $0\nu 2
  \beta$ decays. In grey we again denote points already excluded by
  other (non-cosmological) bounds.}\label{fig:3+1:0nu2beta:m4:nh:ih} 
\end{figure}

\noindent
As previously mentioned, we also revisit the prospects of the
``3+1'' effective model regarding $0\nu 2 \beta$
(constraints on the parameter space, as well as 
the potential for a detection in the near future). In
Fig.~\ref{fig:3+1:0nu2beta:m4:nh:ih} we present the expected ranges
for  $|m_{ee}|$ as a function of the sterile mass. The horizontal
lines denote current bounds and future sensitivities, in agreement with
the discussion in Table~\ref{tab:nulesssensitivities}.
More precisely, we have considered the current bound of 
 $|m_{ee} | \lsim 300 \,\rm
meV$~\cite{Agostini:2013mzu} also showing a future expected sensitivity
of $|m_{ee}| \lsim 100$ meV. 

As can be seen from Fig.~\ref{fig:3+1:0nu2beta:m4:nh:ih}, both
hierarchies can lead to significant values of $|m_{ee}|$, and this
observable plays in fact an important r\^ole in excluding sizeable
regions of the ``3+1'' effective model parameter space. For the heavy mass
regime, a future observation would allow to probe both hierarchies,
and would  correspond to regions of the parameter space compatible
with cosmological constraints. For the low mass regime ($m_4 \lesssim
0.1$ GeV), future experiments can only probe regions 
that would require considering a non-standard cosmology. Notice
however, that for the IH case (which, as expected, is associated to larger
values of $|m_{ee}|$), the low mass regime could be potentially probed
by, for instance SNO in its ``Phase 2''~\cite{Hartnell:2012qd} 
(we did not explicitly include the corresponding future sensitivity in the
plots of  Fig.~\ref{fig:3+1:0nu2beta:m4:nh:ih}). 

It is worth emphasizing here that, under the assumption that
sterile neutrinos are present, a signal in $0 \nu 2 \beta$ 
decay future experiments does not necessarily imply an IH
for the light neutrino spectrum.

Leading to both panels of Fig.~\ref{fig:3+1:0nu2beta:m4:nh:ih}, we
have taken into account non-vanishing values of all (Dirac and
Majorana) phases. The CP-conserving (real) limit would translate into
similar plots - the only significant difference being a less disperse
pattern for the points; we thus refrain from including them here.

\begin{figure}[!h]
\begin{center}
\begin{tabular}{cc}
\includegraphics[width=55mm,angle=270]{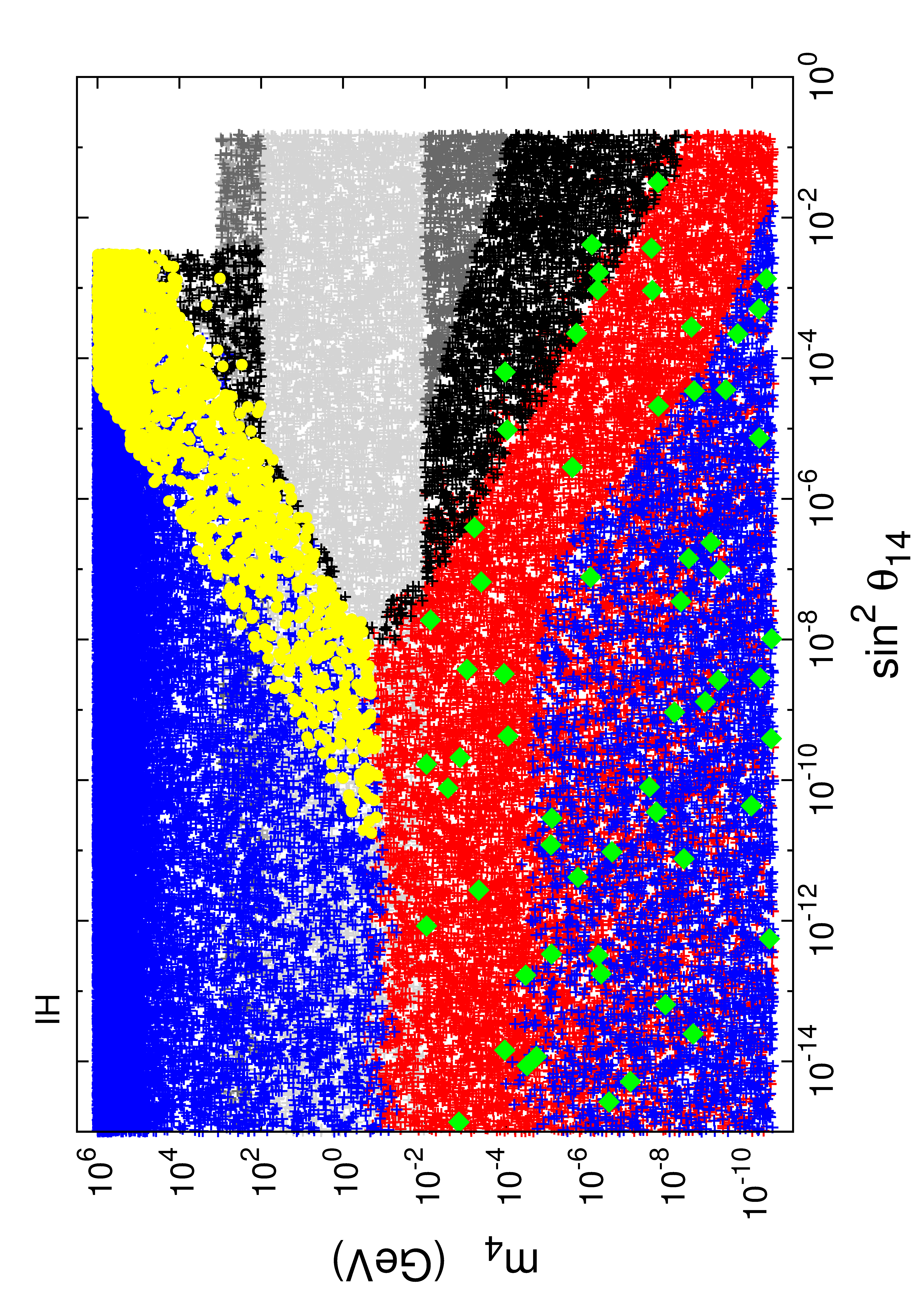}
&
\includegraphics[width=55mm,angle=270]{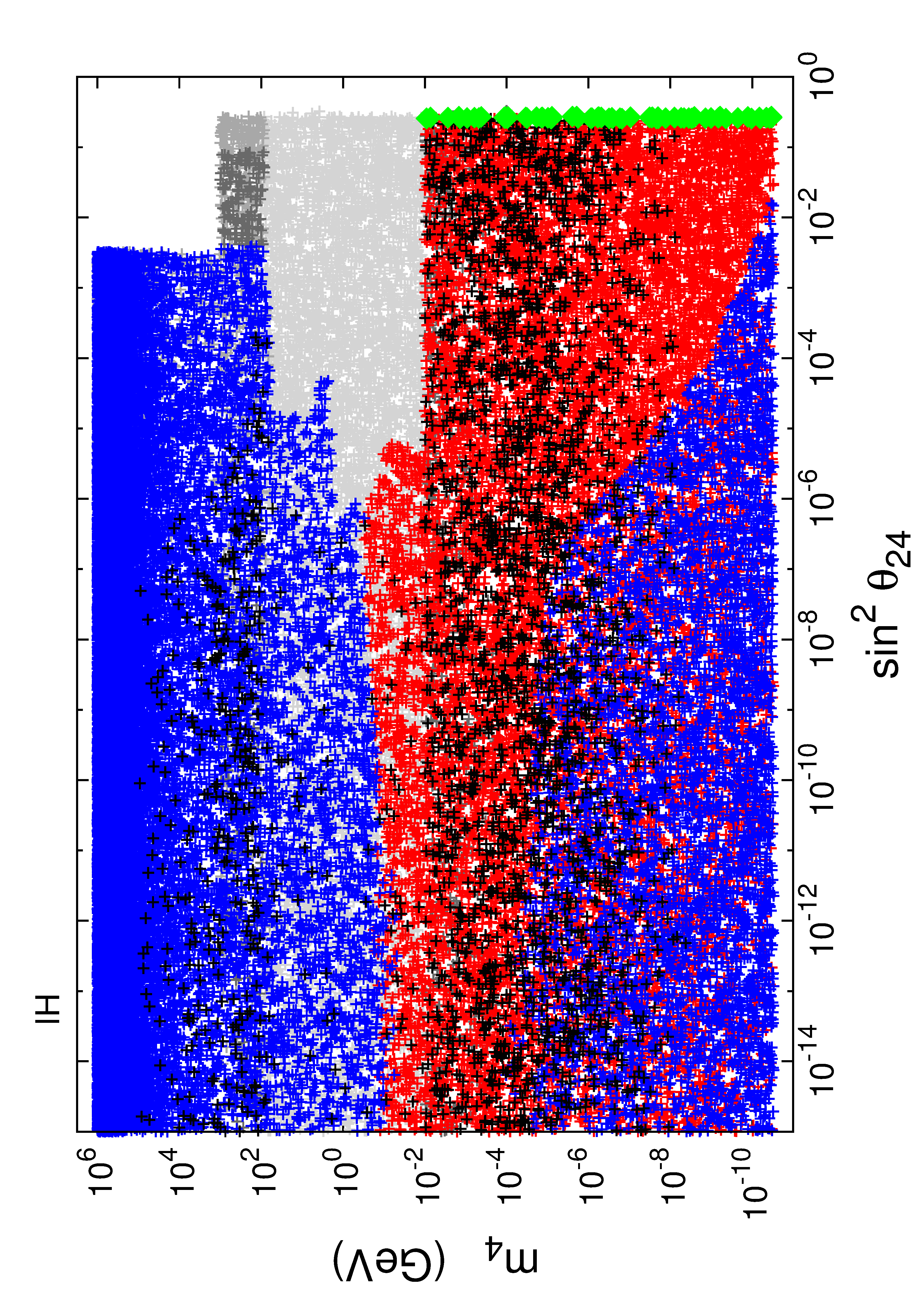}
\end{tabular}
\end{center}
\caption{Effective case: parameter space of the sterile state for an IH light
  neutrino spectrum. On the left, $(\sin^2 \theta_{14}, m_4)$ plane,
  and on the right the $(\sin^2 \theta_{24}, m_4)$ one.  Blue
  points are in agreement with cosmological bounds, while the red ones
  would require considering a non-standard cosmology. Black
regions correspond to violating (at least) the $0 \nu 2 \beta$
constraints, while grey ones are excluded due to conflict with
other (non-cosmological) bounds.  Green diamonds correspond to the 
points associated to $|\Delta (a_\mu)|$ within the
3$\sigma$ interval. 
A yellow band (left panel) further denotes $|m_{ee}|$ within
experimental reach (i.e. $0.05 \text{ eV} \lesssim |m_{ee}|\lesssim 0.1
\text{ eV}$) for cosmologically allowed points.}\label{fig:3+1:m4:sintheta}
\end{figure}

To complete the study conducted within this model, we display in
Fig.~\ref{fig:3+1:m4:sintheta} the ``3+1'' effective parameter space, 
in particular the $(\sin^2 \theta_{i4}, m_4)$ planes. We highlight in black
regions in which (at least) the $0 \nu 2 \beta$ constraints are
violated; those in yellow correspond to cosmologically viable points 
having $|m_{ee}|$ within
experimental reach (i.e. $0.05 \text{ eV} \lesssim |m_{ee}|\lesssim 0.1
\text{ eV}$). Green diamonds correspond to the 
points associated to $|\Delta (a_\mu)|$ within the
3$\sigma$ interval (although some of these points appear to correspond
to regions in agreement with cosmological bounds, we stress that they
correspond to cosmologically disfavoured regimes, as explicitly
displayed in Fig.~\ref{fig:3+1:deltaamu:nh:eta:m4}). 

Figure~\ref{fig:3+1:m4:sintheta} clearly encodes the
phenomenological r\^ole of the active-sterile mixing angles: it
confirms that the largest contributions to $|\Delta (a_\mu)|$
are, as expected, related to large values\footnote{Notice that in this
plot values of $\sin^2 \theta_{14} \simeq 1$, which are excluded due to
conflict with $\nu$-oscillation data, are not displayed.} of 
$\theta_{24}$, while $\theta_{14}$ is the relevant 
parameter regarding $0 \nu 2 \beta$ decays. 
As already commented when discussing
Fig.~\ref{fig:3+1:deltaamu:nh:eta:m4}, 
the largest values for $|\Delta (a_\mu)|$ correspond to regimes of
sizeable sterile-active mixing (as seen from the right panel), 
pulling the entries of the $\tilde
U_\text{PMNS}$ towards the limits of phenomenological and experimental
viability. 
(The analogous analysis of the $(\sin^2 \theta_{34},m_4)$ plane does
not provide any new information, and so we do not display it here.)
The case of a normal hierarchy in the light neutrino spectrum would
lead to similar results for the different $(\sin^2 \theta_{i4}, m_4)$
planes - the exception being the prospects for $0\nu 2 \beta$ decays
(see Fig.~\ref{fig:3+1:0nu2beta:m4:nh:ih}); 
in particular the yellow band in the $(\sin^2
\theta_{14}, m_4)$ plane would be significantly narrower.

\bigskip
\noindent
{\bf Summary for the ``3+1" effective model}

\begin{figure}[!h]
\begin{center}
\begin{tabular}{cc}
\includegraphics[width=55mm,angle=270]{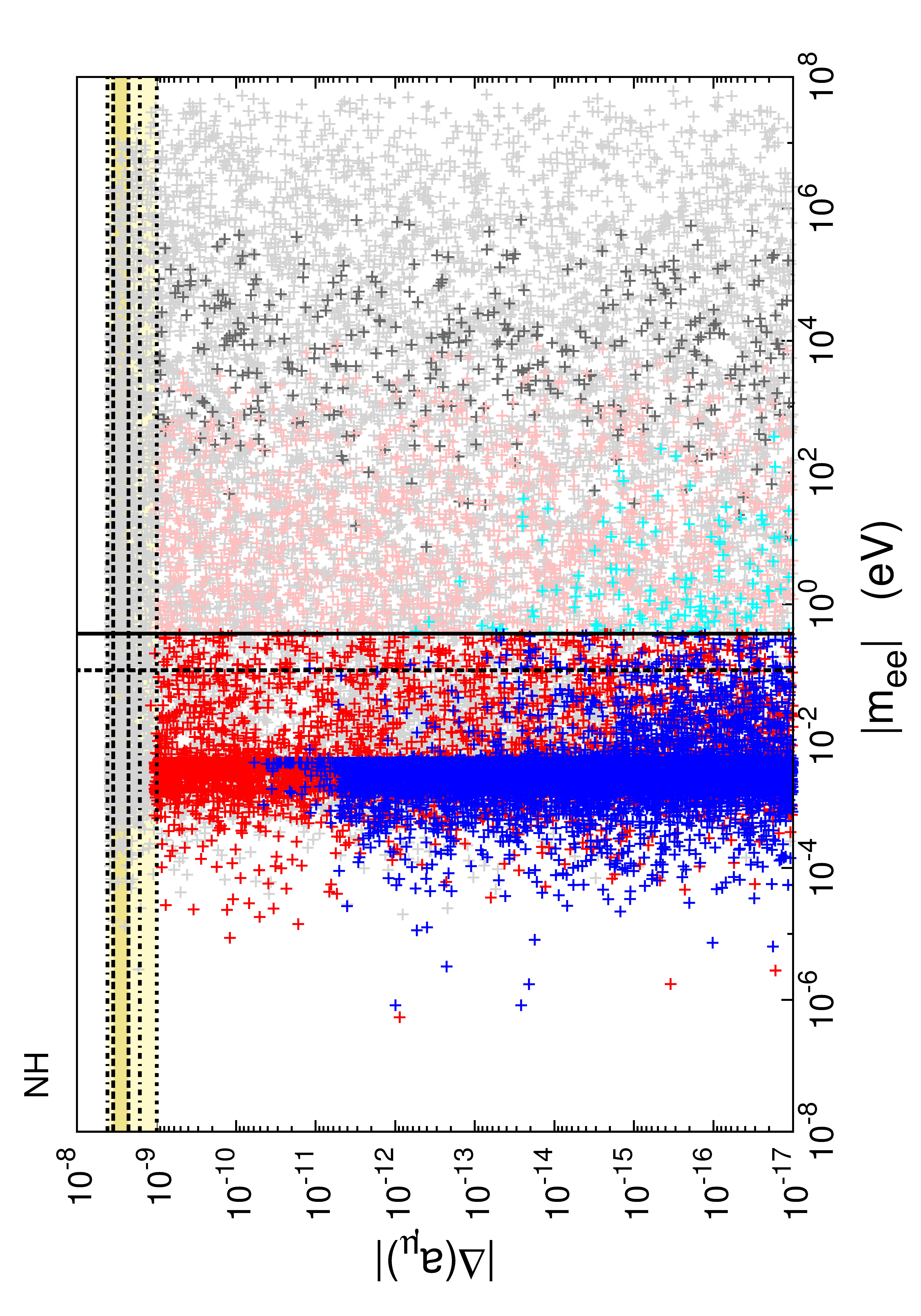}
&
\includegraphics[width=55mm,angle=270]{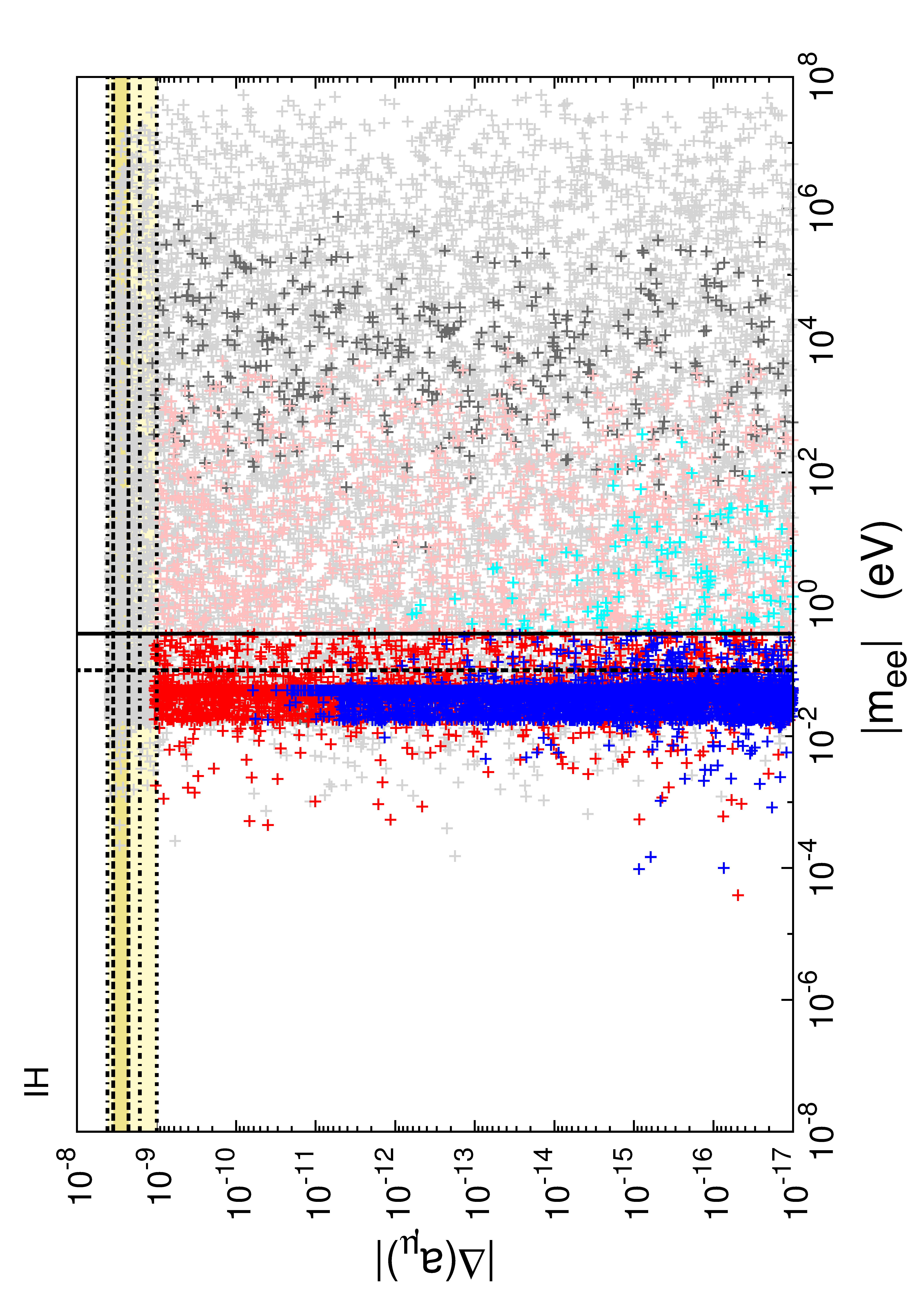}
\end{tabular}
\end{center}
\caption{Effective case: summary of the ``3+1" effective model prospects regarding
  its contributions to $|\Delta (a_\mu)|$ and $0 \nu 2 \beta$ decay.
On the left (right) panel, NH (IH) for the light neutrino spectrum. 
The horizontal lines denote the 1$\sigma$-3$\sigma$ intervals for  
$|\Delta (a_\mu)|$, while vertical full (dashed) correspond to the
current bounds (future sensitivity) for $|m_{ee}|$. 
Colour code as in
Fig.~\ref{fig:3+1:0nu2beta:m4:nh:ih}.}\label{fig:3+1:amu:mee:nh:ih} 
\end{figure}

\noindent
A global overview of the prospects of the 
``3+1" effective model regarding {\it both} $(g-2)_\mu$ and
neutrinoless double beta decay is presented in
Fig.~\ref{fig:3+1:amu:mee:nh:ih}, for NH and IH light neutrino spectra.
Since both observables have already been extensively discussed, we
only stress that in the framework of this simple extension of the SM,
having $|\Delta (a_\mu)|$ within the 3$\sigma$ interval and a possible
observation of $0 \nu 2 \beta$ decay in the next generation of
dedicated facilities requires invoking a non-standard cosmology.

\subsection{Results: Inverse Seesaw scenario}\label{subsec:ISSres}

In the ISS, the numerical contributions to the studied observables are derived
through the following general scan: leading to the construction of the
$9\times 9$ mass matrix in Eq.~(\ref{eq:ISS:M9}),
the moduli of the entries of the matrices $M_R$ and $\mu_X$ are
randomly taken to lie on the intervals
$0.1 \text{ MeV} \lesssim (M_R)_{i}  \lesssim 10^6 \text{ GeV}$ and
$0.01 \text{ eV} \lesssim
(\mu_X)_{ij}  \lesssim 1 \text{ MeV}$, with complex entries for the 
lepton number violating matrix $\mu_X$; 
we also take complex angles for the arbitrary $R$ matrix, randomly 
varying their values in the interval $[0, 2\pi]$.
The modified Casas-Ibarra parametrization for $Y^\nu$,
Eq.~(\ref{eq:YvcasasI}), ensures that constraints from 
neutrino oscillation data are 
satisfied. We use as input the best-fit values of the global analysis
of~\cite{Tortola:2012te}, for each light neutrino hierarchy: NH and IH.

\bigskip
\noindent
{\bf Anomalous magnetic moment of the electron}

\begin{figure}[!h]
\begin{center}
\includegraphics[width=60mm,angle= 270]{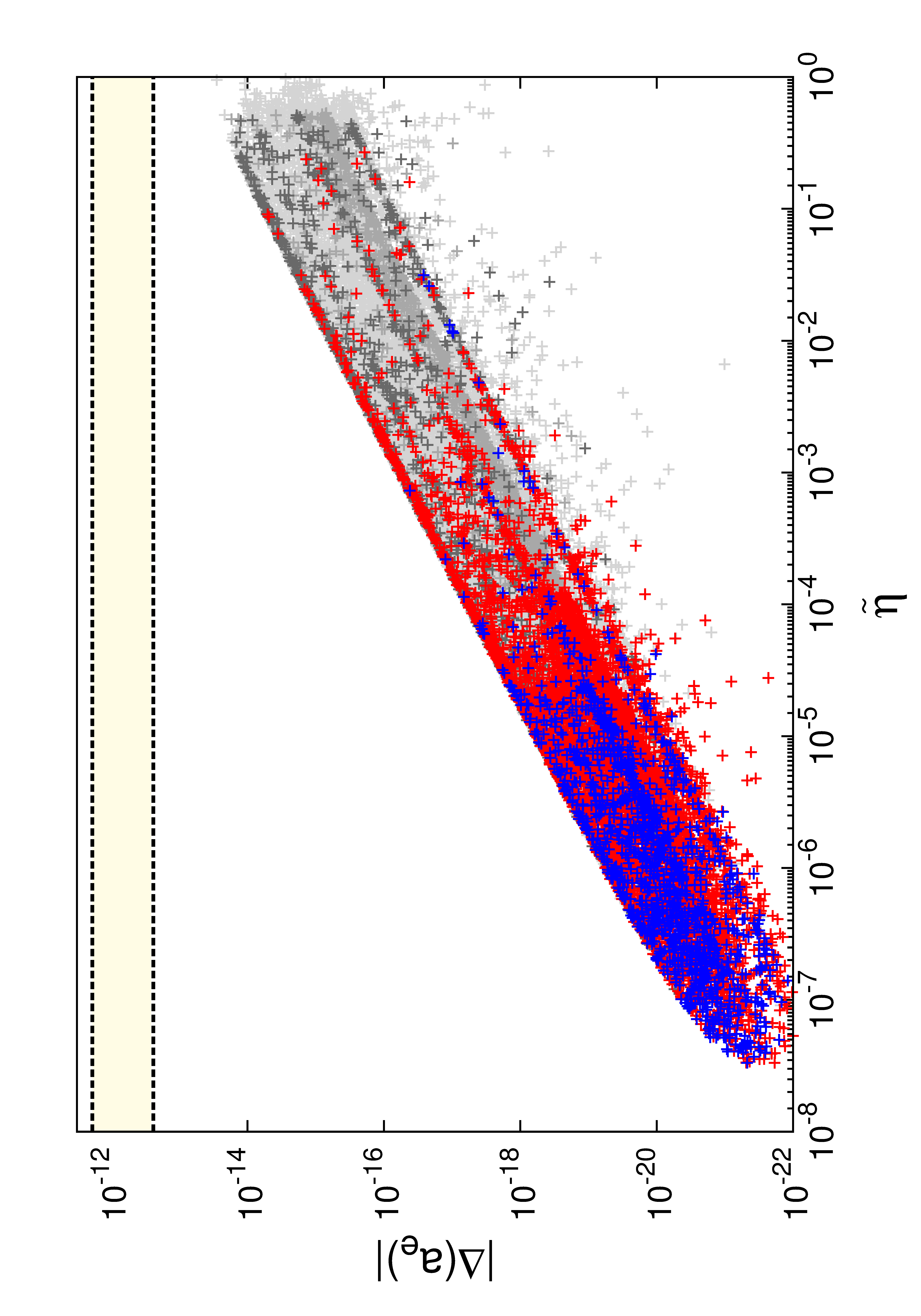}
\end{center}
\caption{ISS: anomalous magnetic moment of the electron as
  a function of $\tilde \eta$, for a NH light neutrino spectrum. 
  Line and colour code as in Fig.~\ref{fig:3+1:deltaae:nh} 
  (see text for a description of the grey-shade scheme).}\label{fig:ISS:deltaae:nh}
\end{figure}

\noindent
As done for the ``3+1'' effective model, we begin the analysis of
the ISS case by investigating the possible constraints arising from
$|\Delta (a_e)|$. Similarly we find that despite having the new 
contributions to $|\Delta (a_e)|$ steadily augmenting
with $\tilde \eta$, the ISS parameter space is not constrained by the
precise determination of this observable. This is illustrated for the
case of a NH in the light neutrino spectrum in Fig.~\ref{fig:ISS:deltaae:nh}.
(The predictions for the IH case are similar.)

Figure~\ref{fig:ISS:deltaae:nh} also provides a first illustration of
some relevant features of the ISS parameter space\footnote{The
comparative smaller dispersion of the points in the ISS case is 
merely due to having an underlying theoretical framework relating the
active, right-handed and sterile neutrinos, so that the different masses
and mixings are not independent degrees of freedom.}: regimes
corresponding to a significant deviation from unitarity of the $\tilde
U_\text{PMNS}$ matrix (i.e. large $\tilde \eta$) are subject to strong
experimental constraints, which exclude important parts of the
parameter space. The grey-shades in Fig.~\ref{fig:ISS:deltaae:nh}
correspond to, at least: failure to comply with $\nu$-oscillation data
(light grey);  violation of unitarity constraints,
bounds from EW precision data, laboratory and LHC bounds, rare decays
such as leptonic meson decays or radiative $\mu \to e \gamma$ decays 
(grey); neutrinoless double beta decays or invisible $Z$-boson width
(dark grey). In particular $\nu$-oscillation data and rare kaon
decay bounds ($\Delta r_K$) are the most relevant ones for the ISS
parameter space.
Finally, and with the exception of a few (isolated) points, notice
that compatibility with cosmological constraints is in general 
obtained only for the regime of small $\tilde \eta$.

\bigskip
\noindent
{\bf Anomalous magnetic moment of the muon}

\begin{figure}[!h]
\begin{center}
\begin{tabular}{cc}
\includegraphics[width=55mm,angle= 270]{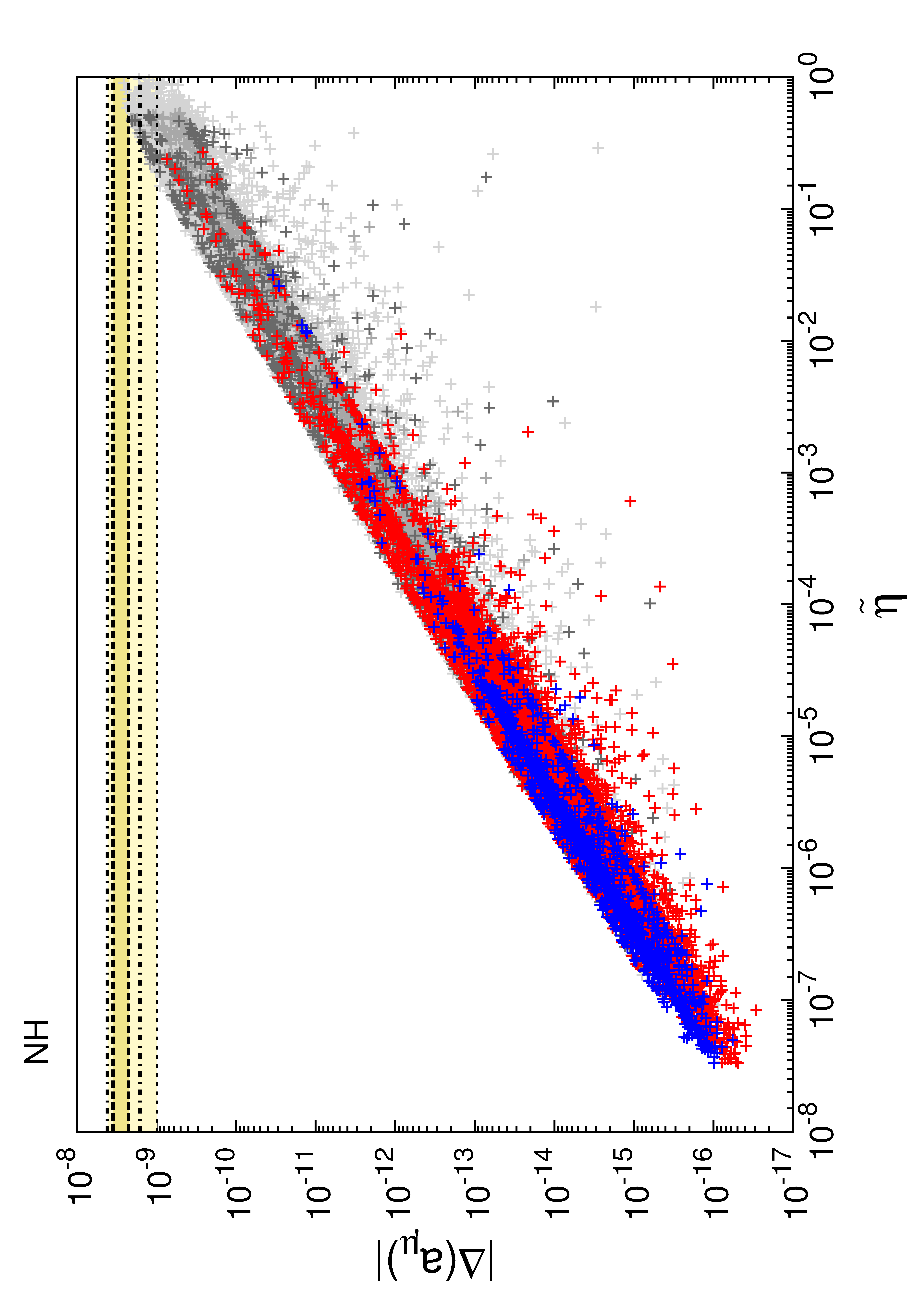}
&
\includegraphics[width=55mm,angle= 270]{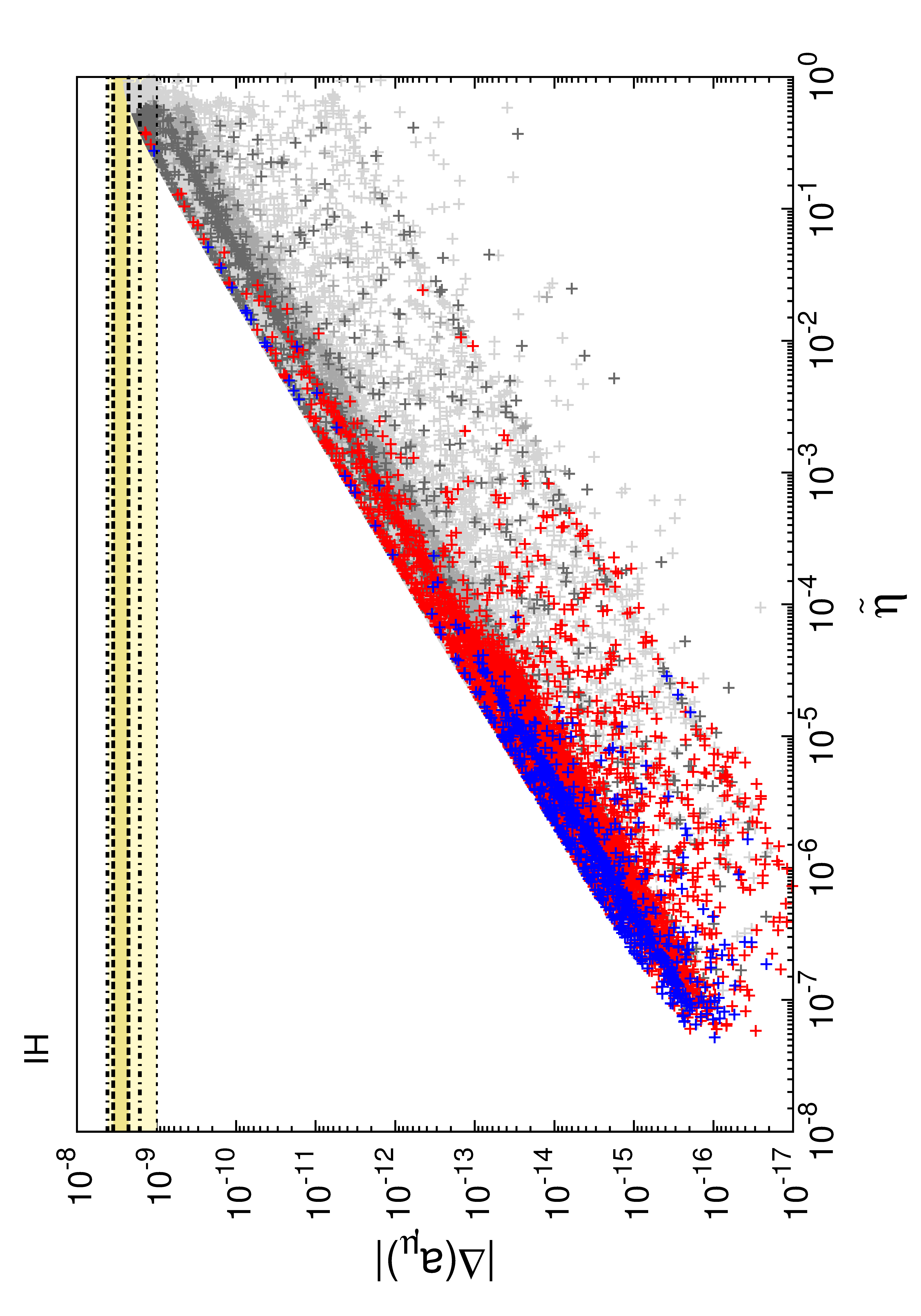}
\end{tabular}
\end{center}
\caption{ISS case: anomalous magnetic moment of the muon as
  a function of $\tilde \eta$ for a NH (left) and an IH (right) 
  light neutrino spectrum. Colour scheme as in
  Fig.~\ref{fig:3+1:deltaamu:nh:eta:m4} (see text for a description of the grey-shade scheme).}\label{fig:ISS:deltaamu:eta:nh:ih}
\end{figure}

\noindent
The contribution of the ISS concerning the anomalous magnetic moment
of the muon is  illustrated, for both light neutrino spectrum
hierarchies, in Fig.~\ref{fig:ISS:deltaamu:eta:nh:ih}. 
As can be seen, only a very small fraction of the points succeeds in
providing a contribution within the 3$\sigma$ interval for $|\Delta
(a_\mu)|$, and only for the IH case (interestingly, a few compatible
with cosmological bounds). We notice that a much larger set would
be within the 3$\sigma$ (even the 2$\sigma$) interval, for both
hierarchies, but these have been excluded as they lead to excessively
large values of $\Delta r_K$ (see Eq.~(\ref{eq:deltarK:value})).

\begin{figure}[!h]
\begin{center}
\begin{tabular}{cc}
\includegraphics[width=65mm,angle= 270]{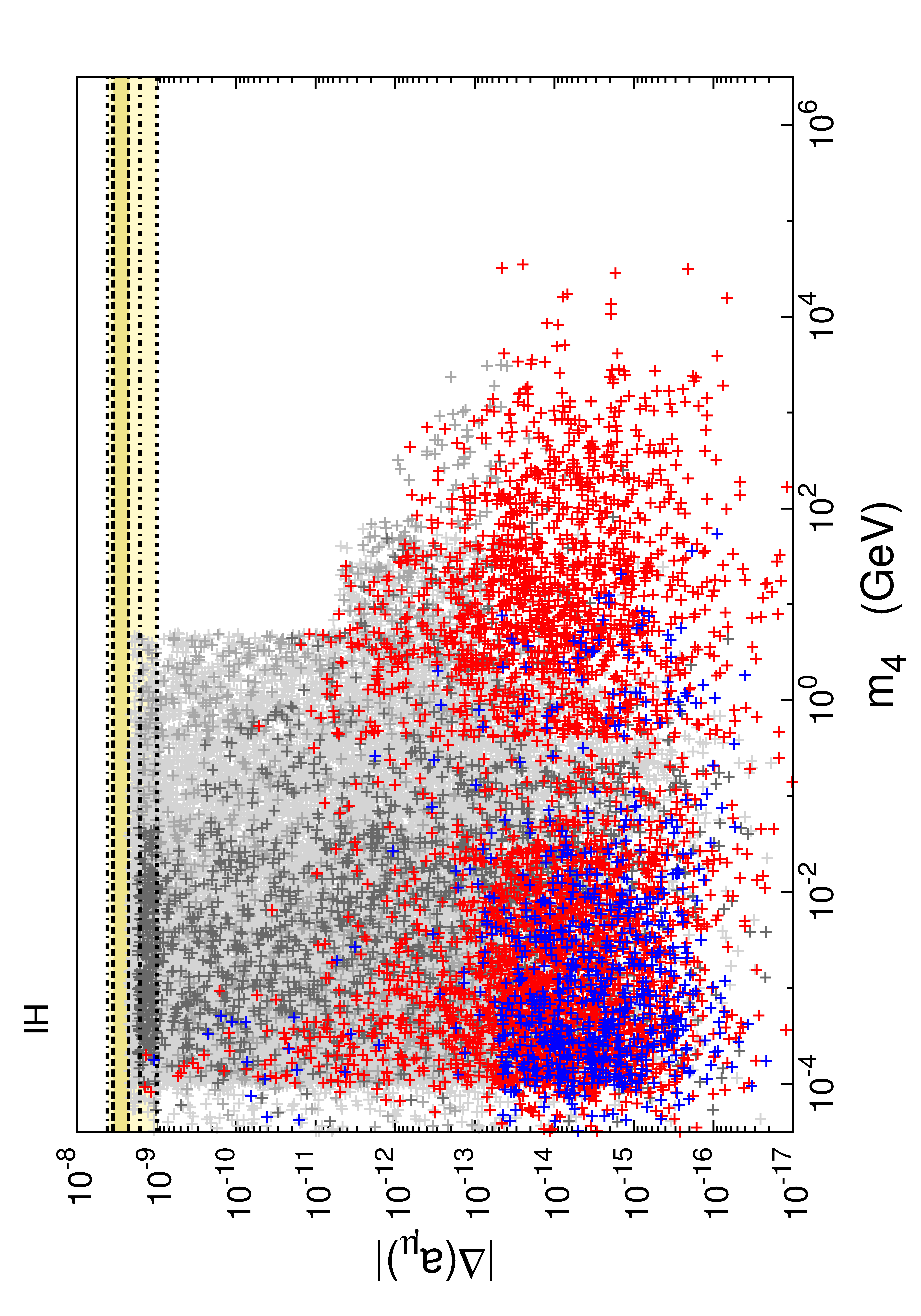}
\end{tabular}
\caption{ISS: anomalous magnetic moment of the muon as
  a function of the mass of the lightest, mostly sterile state ($m_4$),  
  for an IH light neutrino spectrum. Colour scheme as in
  Fig.~\ref{fig:3+1:deltaamu:nh:eta:m4}.}\label{fig:ISS:deltaamu:m4:ih}
\end{center}
\end{figure}

We display in Fig.~\ref{fig:ISS:deltaamu:m4:ih} the 
ISS contributions to the anomalous
magnetic moment of the muon as a function of the lightest
(mostly) sterile state mass\footnote{Contrary to the ``3+1'' effective model analysis, here we do not include the vertical lines, as this realisation cannot address the corresponding observations. However, a different ISS realisation has been recently shown capable of accounting for the $\sim 3.5$ keV line in the X-ray spectra of galaxy clusters \cite{Abada:2014zra}.}, $m_4$, as the lighter states provide the
dominant contribution to this observable (confirmed from
Eq.~(\ref{eq:amu:sterile})). As can be inferred from this
figure, accommodating at the 3$\sigma$ level the discrepancy in
$(g-2)_\mu$ requires that at least one pseudo-Dirac state (pair) be
very light, with a mass around the 0.1~MeV.

%\bigskip
%\bigskip
\newpage
\noindent

{\bf Anomalous magnetic moment of the tau}

\noindent
Such as it occurred for the ``3+1'' effective model, the ISS can only
account for an extremely small contribution to
the anomalous magnetic moment of the tau lepton, 
\begin{equation}\label{eq:res:ISS:deltaatau}
|\Delta (a_{\tau})| \lesssim 10^{-7}\,,
\end{equation}
also well beyond experimental reach. 

\bigskip
\noindent
{\bf Neutrinoless double beta decay}

\begin{figure}[!h]
\begin{center}
\begin{tabular}{cc}
\includegraphics[width=55mm,angle= 270]{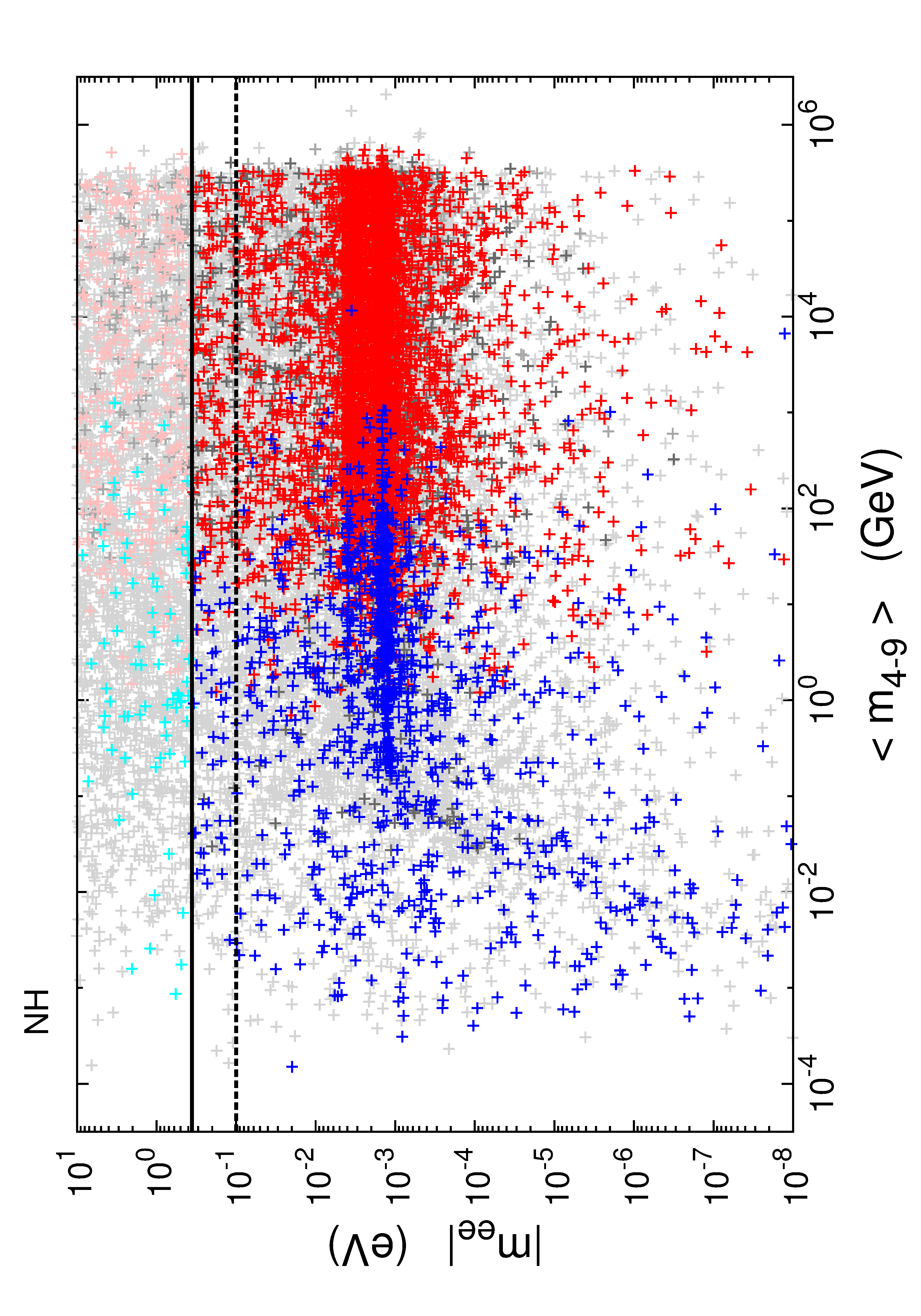}
&
\includegraphics[width=55mm,angle= 270]{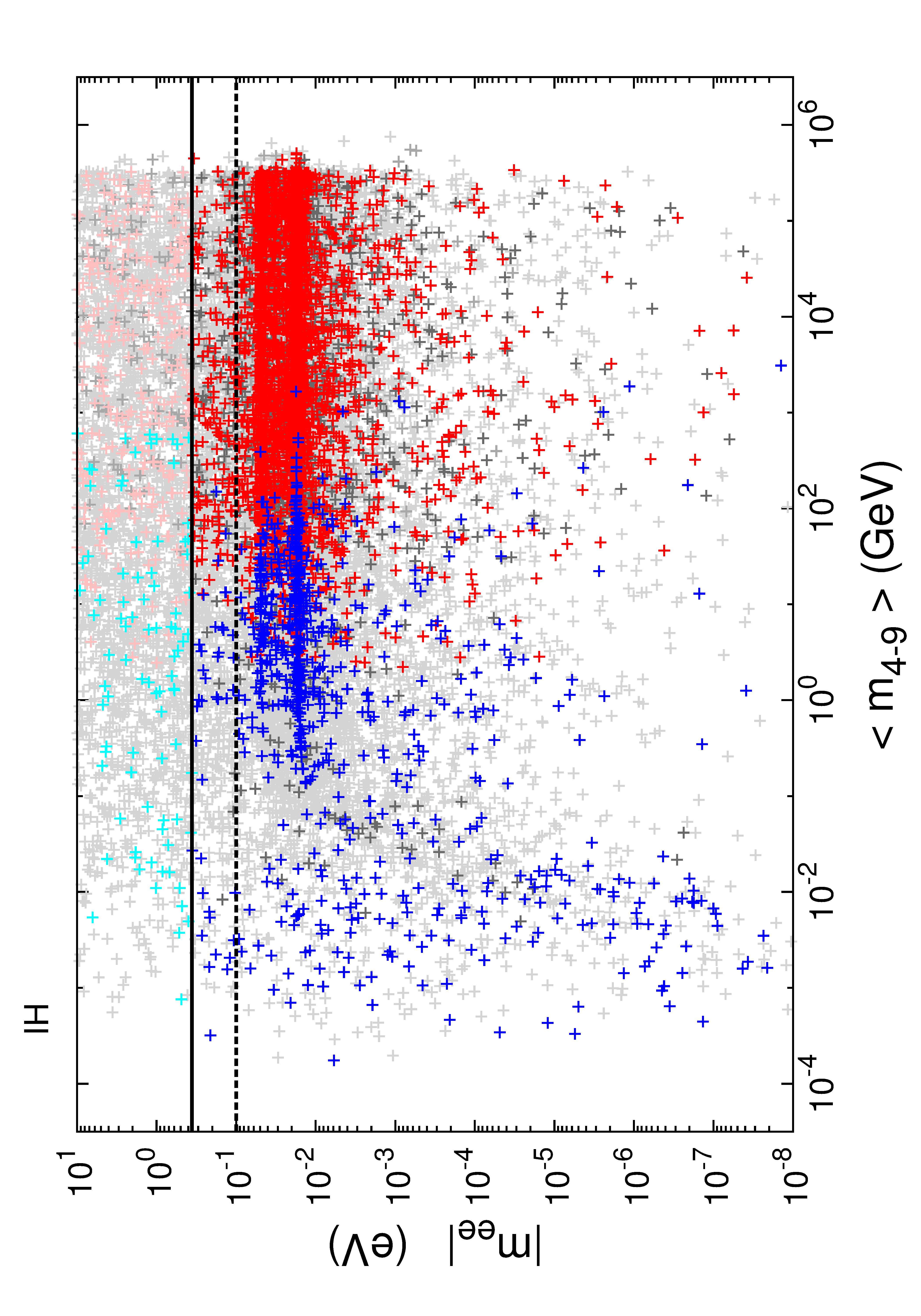}
\end{tabular}
\end{center}
\caption{ISS: effective mass $|m_{ee}|$ (in eV) as a
  function of the average value of the mostly
  sterile state masses, $\langle m_{4-9} \rangle$ 
  (see Eq.~(\ref{eq:ISS:mee:avmass})), for NH (left) and IH
  (right) light neutrino 
  spectra. Line and colour scheme as in
  Fig.~\ref{fig:3+1:0nu2beta:m4:nh:ih}.}\label{fig:ISS:0nu2beta:m4:nh:ih} 
\end{figure}

\noindent
The predictions of the present ISS realisation, with three
right-handed and three sterile neutrino states, are displayed in 
Fig.~\ref{fig:ISS:0nu2beta:m4:nh:ih}, as a function of the average of
the absolute masses of the mostly sterile states, 
\begin{equation}\label{eq:ISS:mee:avmass}
\langle m_{4-9} \rangle\,=\, \sum_{i=4...9} \frac{1}{6}\,|m_i|\,,
\end{equation}
for both hierarchies of the light neutrino spectrum. We take this
average to illustrate our results
as it provides a crude, yet efficient, means to define the 
``heavy'' and ``light'' mass regimes for the sterile states (as a
whole). Leading to both
panels, we have conducted a thorough exploration of the impact of
the CPV phases of the ISS model 
(see scan description in the beginning of this section).

\begin{figure}[!h]
\begin{center}
\begin{tabular}{cc}
\includegraphics[width=60mm,angle= 270]{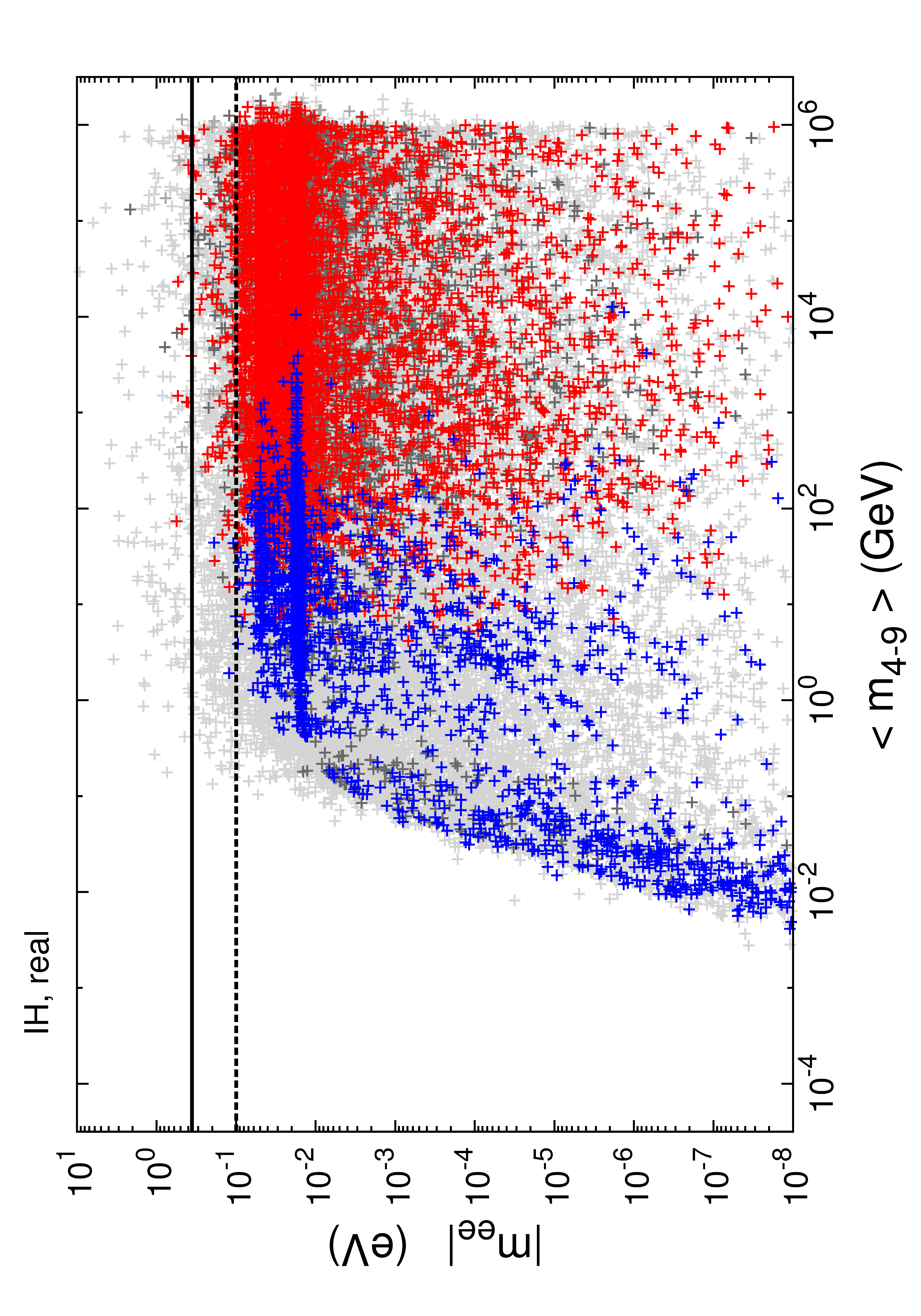}
\end{tabular}
\caption{ISS: effective mass $|m_{ee}|$ (in eV) as a
  function of the average value of the mostly
  sterile state masses, $\langle m_{4-9} \rangle$ 
  (see Eq.~(\ref{eq:ISS:mee:avmass})), for the CP conserving
  limit and an IH light neutrino
  spectrum. Line and colour scheme as in
  Fig.~\ref{fig:3+1:0nu2beta:m4:nh:ih}.}\label{fig:ISS:0nu2beta:m4:real:ih}
\end{center}
\end{figure}

The shape of the ISS contributions to $|m_{ee}|$ is very different
from that encountered in the analysis of the ``3+1" effective
model. This is due to the extended spectrum which, as discussed, is
composed of three pseudo-Dirac pairs, the mass difference in each pair of
$\mathcal{O}(\mu_X)$, and with each element in a pair accounting for
contributions of opposite sign (as can be seen from 
Eq.~(\ref{eq:22bbdecay:ISS})). 
In the absence of CPV phases, these contributions indeed cancel out to
a good approximation, the maximum value saturating around what would be
expected from the SM extended by three light active Majorana neutrinos
(accounting for $\nu$-oscillation data) - this can be inferred from
Fig.~\ref{fig:ISS:0nu2beta:m4:real:ih}, where we plot the CP
conserving case (all Majorana and Dirac phases set to zero) for the IH
case. In this CP conserving limit, one verifies that $0 \nu 2 \beta$
bounds hardly constrain the ISS parameter space, and that a
near future $0 \nu 2 \beta$ signal could only be accounted for by an ISS 
realisation requiring a non-standard cosmology.
Although we do not display it here, the CP conserving NH case could
not account for $|m_{ee}|$ within experimental reach (even for
cosmologically disfavoured points).

The observed behaviour in the above plots has important implications
concerning the interpretation of a possible $0 \nu 2 \beta$ signal in
the near future since, and contrary to other low-scale models of neutrino
mass generation, both hierarchies for the light neutrino spectrum can
account for such an observation; a comparatively light sterile
spectrum, with an average mass scale between $10^{-3}$~GeV and 100~GeV, can account for such a signal, irrespective of the hierarchy, still complying with cosmological observations. Moreover, it would strongly
suggest non-vanishing (Majorana) CPV phases.

\bigskip
\noindent
{\bf Summary for the ``ISS''}

\begin{figure}[!h]
\begin{center}
\begin{tabular}{cc}
\includegraphics[width=55mm,angle=270]{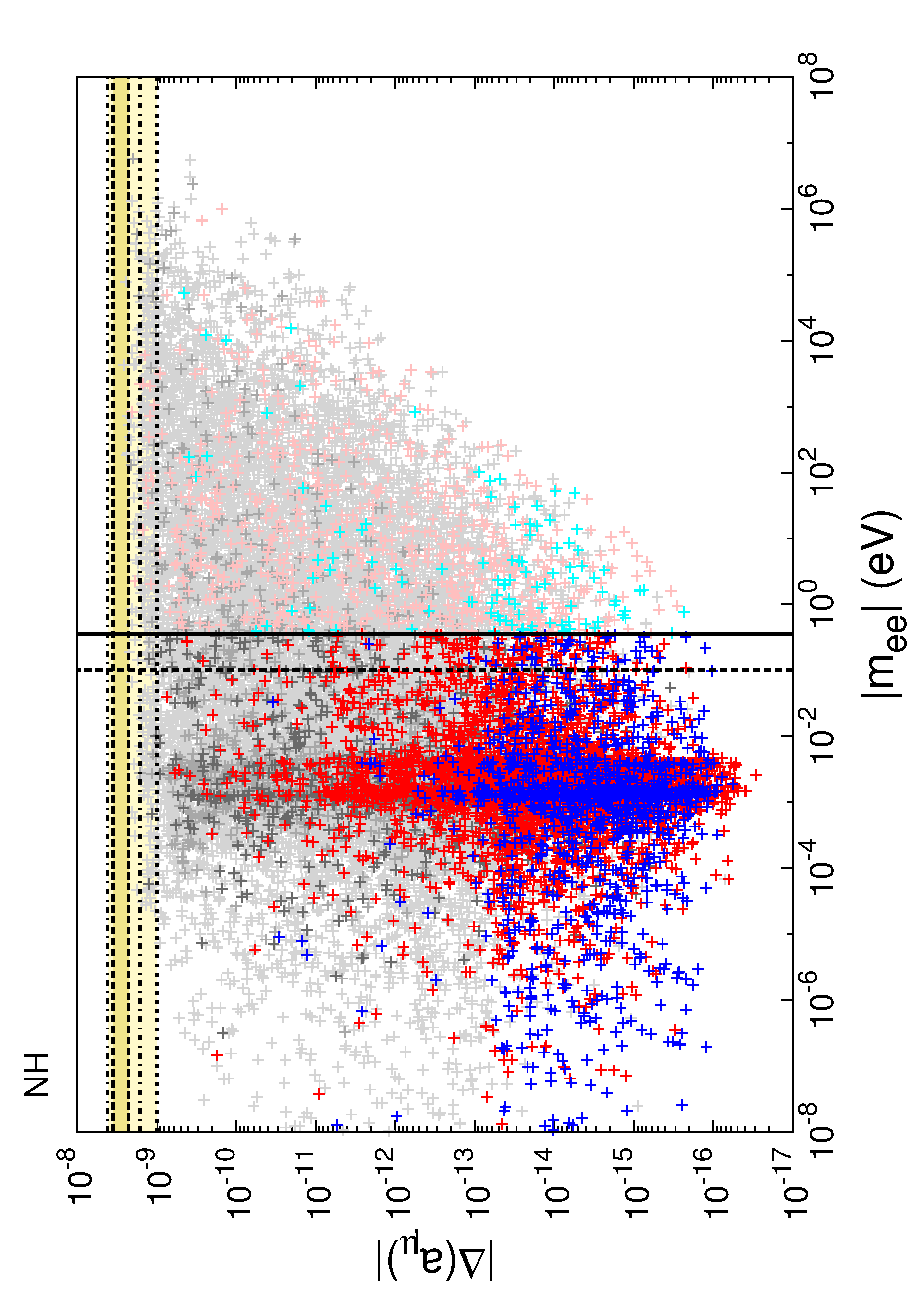}
&
\includegraphics[width=55mm,angle=270]{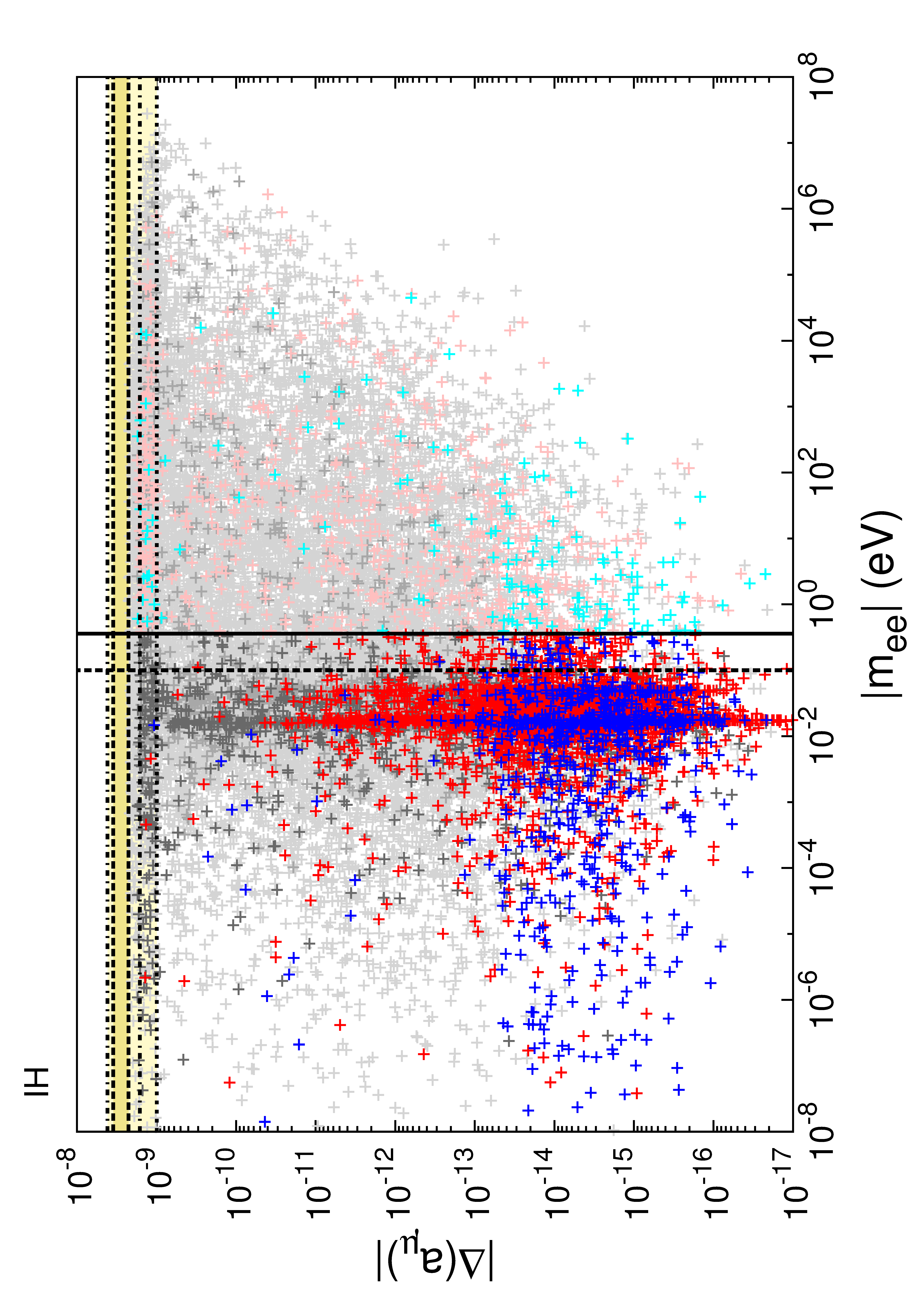}
\end{tabular}
\end{center}
\caption{ISS: overview of the prospects regarding
  the contributions to $|\Delta (a_\mu)|$ and $0 \nu 2 \beta$ decay.
On the left (right) panel, NH (IH) for the light neutrino spectrum. 
The horizontal lines denote the 1$\sigma$-3$\sigma$ intervals for $|\Delta
(a_\mu)|$, while vertical full (dashed) correspond to the current bounds
(future sensitivity) for $|m_{ee}|$. 
Colour code as in Fig.~\ref{fig:3+1:0nu2beta:m4:nh:ih}.}\label{fig:ISS:amu:mee:nh:ih}
\end{figure}

\noindent
As done for the ``3+1" effective model, we summarise the prospects of
the ISS concerning $(g-2)_\mu$ and neutrinoless double beta
decays; an overview, for both NH and IH light neutrino spectra, is
shown in Fig.~\ref{fig:ISS:amu:mee:nh:ih}. One can conclude that
should an ISS mechanism be at the origin of neutrino mass generation, 
a realisation accounting for a near future observation of a
neutrinoless double beta decay signal cannot alleviate the tension in
$(g-2)_\mu$, not even at the 3$\sigma$ level. 

\section{Conclusions}\label{sec:concs}

In this work we investigated the r\^ole of sterile neutrinos on the
(anomalous) magnetic moment of leptons, as well as their contribution
to the neutrinoless double beta decay effective mass.
We considered minimal extensions of the SM by sterile 
fermion states. The simplest case is that of an effective construction where,
without any assumption on the neutrino mass and mixing generation 
mechanism, the neutral fermion spectrum contains an additional massive state, which
mixes with the active states, leading to an enlarged leptonic mixing
matrix. We then focused on a specific well-motivated framework, which
consists in the embedding of an Inverse Seesaw mechanism into the SM,
extended by three right-handed neutrinos and three sterile states. 

Our study reveals that the simple``3+1" effective
extension of the SM can account for a contribution capable of
alleviating the tension between theory and experiment on $(g-2)_\mu$,
for large values of $\tilde
\eta$, and a light sterile state, $m_4 \lesssim 10^{-2}$ GeV
(although only for points disfavoured by cosmological observations). 
Regarding $0 \nu 2 \beta$ decays, and in the regime of a heavy sterile
mass, a future observation would allow to probe both hierarchies
(for a spectrum compatible with cosmological constraints). 
For the low mass regime ($m_4 \lesssim 0.1$ GeV), future $0 \nu 2 \beta$ 
experiments can only probe regions of the parameter
space that would require considering a non-standard cosmology. 
As shown in our analysis, and under the assumption that
sterile neutrinos are present, a  signal in $0 \nu 2 \beta$ 
decay future experiments does not necessarily imply an IH
for the light neutrino spectrum. Finally, we have shown that 
having $|\Delta (a_\mu)|$ within the 3$\sigma$ interval and a possible
observation of $0 \nu 2 \beta$ decay in the next generation of
dedicated experiments, requires invoking a non-standard cosmology.

The analysis of the ISS scenario revealed 
that a contribution within the 3$\sigma$ interval for $|\Delta
(a_\mu)|$ is indeed possible, albeit corresponding to a very small
fraction of the phenomenological viable ISS parameter space (and only
for the IH case), as the most promising regions are 
excluded due to excessively large values of 
$\Delta r_K$. Regarding  $|m_{ee}|$, and 
contrary to other low-scale models of neutrino
mass generation, in the considered ISS configuration 
both hierarchies (IH and NH) can
account for an observation in near future facilities. 
We have also argued that should an ISS 
mechanism be indeed at the origin of neutrino mass generation, 
a realisation accounting for a near future observation of a
neutrinoless double beta decay signal cannot alleviate the tension in
$(g-2)_\mu$, not even at the 3$\sigma$ level. 

Finally, the contribution to the electron and tau magnetic moments lies in both
cases beyond experimental reach.

\section*{Acknowledgements}
The authors acknowledge support from the European Union FP7 ITN
INVISIBLES (Marie Curie Actions, PITN-\-GA-\-2011-\-289442). 
V.D.R. and A.M.T. are grateful for 
the kind hospitality of the LPT-Orsay, where part of this work 
was carried out.


\begin{thebibliography}{99}


 \bibitem{reactor:I}
%\bibitem{Mueller:2011nm}
  T.~A.~Mueller, D.~Lhuillier, M.~Fallot, A.~Letourneau, S.~Cormon,
  M.~Fechner, L.~Giot and T.~Lasserre {\it et al.}, 
  %``Improved Predictions of Reactor Antineutrino Spectra,''
  Phys.\ Rev.\ C {\bf 83} (2011) 054615
  [arXiv:1101.2663 [hep-ex]];
%\bibitem{Huber:2011wv}
  P.~Huber,
  %``On the determination of anti-neutrino spectra from nuclear reactors,''
  Phys.\ Rev.\ C {\bf 84} (2011) 024617
   [Erratum-ibid.\ C {\bf 85} (2012) 029901]
  [arXiv:1106.0687 [hep-ph]];
%\bibitem{Mention:2011rk}
  G.~Mention, M.~Fechner, T.~Lasserre, T.~A.~Mueller, D.~Lhuillier,
  M.~Cribier and A.~Letourneau, 
  %``The Reactor Antineutrino Anomaly,''
  Phys.\ Rev.\ D {\bf 83} (2011) 073006
  [arXiv:1101.2755 [hep-ex]].
  
\bibitem{Aguilar:2001ty}
  A.~A.~Aguilar-Arevalo {\it et al.}  [LSND Collaboration], 
 % \emph{Evidence for neutrino oscillations from the observation of
 % anti-neutrino(electron) appearance in a anti-neutrino(muon) beam}, 
Phys.\ Rev.\ D {\bf 64} (2001) 112007
  [hep-ex/0104049].
 
 \bibitem{miniboone:I} 
%\bibitem{AguilarArevalo:2007it}
  A.~A.~Aguilar-Arevalo {\it et al.}  [MiniBooNE Collaboration],
  %``A Search for electron neutrino appearance at the $\Delta m^{2} \sim 1$eV$^{2}$ scale,''
  Phys.\ Rev.\ Lett.\  {\bf 98} (2007) 231801
  [arXiv:0704.1500 [hep-ex]];
%\bibitem{AguilarArevalo:2010wv} 
  A.~A.~Aguilar-Arevalo {\it et al.}  [MiniBooNE Collaboration],
  %``Event Excess in the MiniBooNE Search for $\bar \nu_\mu
  %\rightarrow \bar \nu_e$ Oscillations,'' 
  Phys.\ Rev.\ Lett.\  {\bf 105} (2010) 181801
  [arXiv:1007.1150 [hep-ex]];
%\bibitem{Aguilar-Arevalo:2013pmq}
  A.~A.~Aguilar-Arevalo {\it et al.}  [MiniBooNE Collaboration],
  %``Improved Search for $\bar \nu_\mu \rightarrow \bar \nu_e$
  %Oscillations in the MiniBooNE Experiment,'' 
  Phys.\ Rev.\ Lett.\  {\bf 110} (2013) 161801
  [arXiv:1207.4809 [hep-ex], arXiv:1303.2588 [hep-ex]].
  
  \bibitem{gallium:I}
%\bibitem{Acero:2007su}
  M.~A.~Acero, C.~Giunti and M.~Laveder,
  %``Limits on nu(e) and anti-nu(e) disappearance from Gallium and reactor experiments,''
  Phys.\ Rev.\ D {\bf 78} (2008) 073009
  [arXiv:0711.4222 [hep-ph]];
%\bibitem{Giunti:2010zu}
  C.~Giunti and M.~Laveder,
  %``Statistical Significance of the Gallium Anomaly,''
  Phys.\ Rev.\ C {\bf 83} (2011) 065504
  [arXiv:1006.3244 [hep-ph]].
  
\bibitem{Kusenko:2009up}
  A.~Kusenko,
  %``Sterile neutrinos: The Dark side of the light fermions,''
  Phys.\ Rept.\  {\bf 481} (2009) 1
  [arXiv:0906.2968 [hep-ph]].
  %%CITATION = ARXIV:0906.2968;%%

\bibitem{Abazajian:2012ys}
  K.~N.~Abazajian, M.~A.~Acero, S.~K.~Agarwalla,
  A.~A.~Aguilar-Arevalo, C.~H.~Albright, S.~Antusch, C.~A.~Arguelles
  and A.~B.~Balantekin {\it et al.}, 
  ``Light Sterile Neutrinos: A White Paper,''
  arXiv:1204.5379 [hep-ph].
  %%CITATION = ARXIV:1204.5379;%%

\bibitem{Schechter:1980gr}
  J.~Schechter and J.~W.~F.~Valle,
  %``Neutrino Masses In SU(2) X U(1) Theories,''
  Phys.\ Rev.\ D {\bf 22} (1980) 2227.
  %%CITATION = PHRVA,D22,2227;%%

\bibitem{Gronau:1984ct}
  M.~Gronau, C.~N.~Leung and J.~L.~Rosner,
  %``Extending Limits on Neutral Heavy Leptons,''
  Phys.\ Rev.\ D {\bf 29} (1984) 2539.
  %%CITATION = PHRVA,D29,2539;%% 

\bibitem{Ilakovac:1994kj}
  A.~Ilakovac and A.~Pilaftsis,
  %``Flavor violating charged lepton decays in seesaw-type models,''
  Nucl.\ Phys.\ B {\bf 437} (1995) 491
  [hep-ph/9403398].
  %%CITATION = HEP-PH/9403398;%%

\bibitem{Deppisch:2004fa}
  F.~Deppisch and J.~W.~F.~Valle,
  %``Enhanced lepton flavor violation in the supersymmetric inverse seesaw model,''
  Phys.\ Rev.\ D {\bf 72} (2005) 036001
  [hep-ph/0406040].
  %%CITATION = HEP-PH/0406040;%%

\bibitem{Arganda:2014via} 
  E.~Arganda, M.~J.~Herrero, X.~Marcano and C.~Weiland,
  ``Lepton flavour violating Higgs decays,''
  arXiv:1406.0384 [hep-ph].

\bibitem{Shrock}
%\bibitem{Shrock:1980vy}
  R.~E.~Shrock,
  %``New Tests For, and Bounds On, Neutrino Masses and Lepton Mixing,''
  Phys.\ Lett.\ B {\bf 96} (1980) 159; 
  %%CITATION = PHLTA,B96,159;%%
%\bibitem{Shrock:1980ct}
  R.~E.~Shrock,
  %``General Theory of Weak Leptonic and Semileptonic
  %Decays. 1. Leptonic Pseudoscalar Meson Decays, with Associated
  %Tests For, and Bounds on, Neutrino Masses and Lepton Mixing,'' 
  Phys.\ Rev.\ D {\bf 24} (1981) 1232.
  %%CITATION = PHRVA,D24,1232;%%

\bibitem{Nardi:1994iv}
  E.~Nardi, E.~Roulet and D.~Tommasini,
  %``Limits on neutrino mixing with new heavy particles,''
  Phys.\ Lett.\ B {\bf 327} (1994) 319
  [hep-ph/9402224].
  %%CITATION = HEP-PH/9402224;%%

\bibitem{Abada:2012mc}
  A.~Abada, D.~Das, A.~M.~Teixeira, A.~Vicente and C.~Weiland,
  %``Tree-level lepton universality violation in the presence of
  %sterile neutrinos: impact for $R_K$ and $R_\pi$,'' 
  JHEP {\bf 1302} (2013) 048
  [arXiv:1211.3052 [hep-ph]].

\bibitem{Abada:2013aba} 
  A.~Abada, A.~M.~Teixeira, A.~Vicente and C.~Weiland,
  %``Sterile neutrinos in leptonic and semileptonic decays,''
  JHEP {\bf 1402} (2014) 091 
  [arXiv:1311.2830 [hep-ph]].

\bibitem{Akhmedov:2013hec}
  E.~Akhmedov, A.~Kartavtsev, M.~Lindner, L.~Michaels and J.~Smirnov,
  %``Improving Electro-Weak Fits with TeV-scale Sterile Neutrinos,''
  JHEP {\bf 1305} (2013) 081
  [arXiv:1302.1872 [hep-ph]].
  %%CITATION = ARXIV:1302.1872;%%

\bibitem{BhupalDev:2012zg}
P.~S.~Bhupal Dev, R.~Franceschini and R.~N.~Mohapatra,
%``Bounds on TeV Seesaw Models from LHC Higgs Data,''
Phys.\ Rev.\ D {\bf 86} (2012) 093010
[arXiv:1207.2756 [hep-ph]].
%%CITATION = ARXIV:1207.2756;%%
 
\bibitem{Das:2012ze}
A.~Das and N.~Okada,
%``Inverse seesaw neutrino signatures at the LHC and ILC,''
Phys.\ Rev.\ D {\bf 88} (2013) 11,  113001
[arXiv:1207.3734 [hep-ph]].
%%CITATION = ARXIV:1207.3734;%%

\bibitem{Cely:2012bz}
C.~G.~Cely, A.~Ibarra, E.~Molinaro and S.~T.~Petcov,
%``Higgs Decays in the Low Scale Type I See-Saw Model,''
Phys.\ Lett.\ B {\bf 718} (2013) 957
[arXiv:1208.3654 [hep-ph]].
%%CITATION = ARXIV:1208.3654;%%
  
\bibitem{Bandyopadhyay:2012px}
  P.~Bandyopadhyay, E.~J.~Chun, H.~Okada and J.~-C.~Park,
  %``Higgs Signatures in Inverse Seesaw Model at the LHC,''
  JHEP {\bf 1301} (2013) 079
  [arXiv:1209.4803 [hep-ph]].
  %%CITATION = ARXIV:1209.4803;%%


\bibitem{Deppisch:2012nb}
  F.~F.~Deppisch, M.~Hirsch and H.~Pas,
  %``Neutrinoless Double Beta Decay and Physics Beyond the Standard Model,''
  J.\ Phys.\ G {\bf 39} (2012) 124007
  [arXiv:1208.0727 [hep-ph]].
  %%CITATION = ARXIV:1208.0727;%%

\bibitem{LHC.LNV}
%\bibitem{Chrzaszcz:2013uz}
  M.~Chrzaszcz,
 ``Searches for LFV and LNV Decays at LHCb,''
  arXiv:1301.2088 [hep-ex].
  %%CITATION = ARXIV:1301.2088;%%

\bibitem{Hanneke:2008tm}
  D.~Hanneke, S.~Fogwell and G.~Gabrielse,
  %``New Measurement of the Electron Magnetic Moment and the Fine Structure Constant,''
  Phys.\ Rev.\ Lett.\  {\bf 100} (2008) 120801
  [arXiv:0801.1134 [physics.atom-ph]].
  %%CITATION = ARXIV:0801.1134;%%

\bibitem{Beringer:1900zz} 
  J.~Beringer {\it et al.}  [Particle Data Group Collaboration],
  %``Review of Particle Physics (RPP),''
  Phys.\ Rev.\ D {\bf 86}  (2012) 010001.

\bibitem{Freitas:2014pua}
  A.~Freitas, J.~Lykken, S.~Kell and S.~Westhoff,
  %``Testing the Muon g-2 Anomaly at the LHC,''
  JHEP {\bf 1405} (2014) 145
  [arXiv:1402.7065 [hep-ph]].
  %%CITATION = ARXIV:1402.7065;%%

\bibitem{Asaka:2005an}
  T.~Asaka, S.~Blanchet and M.~Shaposhnikov,
  %``The nuMSM, dark matter and neutrino masses,''
  Phys.\ Lett.\ B {\bf 631} (2005) 151
  [hep-ph/0503065].
  %%CITATION = HEP-PH/0503065;%%

\bibitem{Ibarra:2010xw}
  A.~Ibarra, E.~Molinaro and S.~T.~Petcov,
  %``TeV Scale See-Saw Mechanisms of Neutrino Mass Generation, the
  %Majorana Nature of the Heavy Singlet Neutrinos and
  %$(\beta\beta)_{0\nu}$-Decay,'' 
  JHEP {\bf 1009} (2010) 108
  [arXiv:1007.2378 [hep-ph]].
  %%CITATION = ARXIV:1007.2378;%%

\bibitem{Mohapatra:1986bd}
  R.~N.~Mohapatra and J.~W.~F.~Valle,
  %``Neutrino Mass and Baryon Number Nonconservation in Superstring Models,''
  Phys.\ Rev.\ D {\bf 34} (1986) 1642.
  %%CITATION = PHRVA,D34,1642;%%

\bibitem{Tortola:2012te} 
D.~V.~Forero, M.~Tortola and J.~W.~F.~Valle,
%``Global status of neutrino oscillation parameters after Neutrino-2012,''
Phys.\ Rev.\ D {\bf 86} (2012) 073012 
[arXiv:1205.4018 [hep-ph]].
  
\bibitem{Fogli:2012ua}
G.~L.~Fogli, E.~Lisi, A.~Marrone, D.~Montanino, A.~Palazzo and A.~M.~Rotunno,
  %``Global analysis of neutrino masses, mixings and phases: entering
  %the era of leptonic CP violation searches,'' 
Phys.\ Rev.\ D {\bf 86} (2012) 013012
[arXiv:1205.5254 [hep-ph]].
 %%CITATION = ARXIV:1205.5254;%%

\bibitem{GonzalezGarcia:2012sz}
M.~C.~Gonzalez-Garcia, M.~Maltoni, J.~Salvado and T.~Schwetz,
%``Global fit to three neutrino mixing: critical look at present precision,''
JHEP {\bf 1212} (2012) 123
[arXiv:1209.3023 [hep-ph]].
%%CITATION = ARXIV:1209.3023;%%

\bibitem{Forero:2014bxa} 
D.~V.~Forero, M.~Tortola and J.~W.~F.~Valle,
``Neutrino oscillations refitted,''
arXiv:1405.7540 [hep-ph].

\bibitem{nufit}
See also http://www.nu-fit.org/

\bibitem{Antusch:2006vwa}
S.~Antusch, C.~Biggio, E.~Fernandez-Martinez, M.~B.~Gavela and J.~Lopez-Pavon,
%``Unitarity of the Leptonic Mixing Matrix,''
JHEP {\bf 0610} (2006) 084 
[arXiv:hep-ph/0607020].

\bibitem{Antusch:2008tz} 
S.~Antusch, J.~P.~Baumann and E.~Fernandez-Martinez,
%``Non-Standard Neutrino Interactions with Matter from Physics Beyond the Standard Model,''
Nucl.\ Phys.\ B {\bf 810} (2009) 369 
[arXiv:0807.1003 [hep-ph]].

\bibitem{Lello:2012gi}
L.~Lello and D.~Boyanovsky,
%``Searching for sterile neutrinos from $\pi$ and $K$ decays,''
Phys.\ Rev.\ D {\bf 87} (2013) 073017
[arXiv:1208.5559 [hep-ph]].
%%CITATION = ARXIV:1208.5559;%%

\bibitem{delAguila:2008pw}
F.~del Aguila, J.~de Blas and M.~Perez-Victoria,
%``Effects of new leptons in Electroweak Precision Data,''
Phys.\ Rev.\ D {\bf 78} (2008) 013010
[arXiv:0803.4008 [hep-ph]].
%%CITATION = ARXIV:0803.4008;%%

\bibitem{Basso:2013jka}
L.~Basso, O.~Fischer and J.~J.~van der Bij,
%``Precision tests of unitarity in leptonic mixing,''
Europhys.\ Lett.\  {\bf 105} (2014) 11001
[arXiv:1310.2057 [hep-ph]].
%%CITATION = ARXIV:1310.2057;%%

\bibitem{Smirnov:2006bu}
A.~Y.~Smirnov and R.~Zukanovich Funchal,
%``Sterile neutrinos: Direct mixing effects versus induced mass matrix of active neutrinos,''
Phys.\ Rev.\ D {\bf 74} (2006) 013001
[hep-ph/0603009].
%%CITATION = HEP-PH/0603009;%%  

\bibitem{Agostini:2013mzu}
M.~Agostini {\it et al.}  [GERDA Collaboration],
%``Results on Neutrinoless Double-$\beta$ Decay of $^{76}$Ge from
%Phase I of the GERDA Experiment,'' 
Phys.\ Rev.\ Lett.\  {\bf 111} (2013) 12,  122503
[arXiv:1307.4720 [nucl-ex]].
%%CITATION = ARXIV:1307.4720;%%

\bibitem{Aoyama:2007dv} 
  T.~Aoyama, M.~Hayakawa, T.~Kinoshita and M.~Nio,
  %``Revised value of the eighth-order electron g-2,''
  Phys.\ Rev.\ Lett.\  {\bf 99} (2007) 110406 
  [arXiv:0706.3496 [hep-ph]].
  %%CITATION = ARXIV:0706.3496;%%

\bibitem{Aoyama:2008hz} 
  T.~Aoyama, M.~Hayakawa, T.~Kinoshita and M.~Nio,
  %``Tenth-Order Lepton Anomalous Magnetic Moment: Second-Order Vertex
  %Containing Two Vacuum Polarization Subdiagrams, One Within the
  %Other,'' 
  Phys.\ Rev.\ D {\bf 78} (2008) 113006 
  [arXiv:0810.5208 [hep-ph]].

\bibitem{Giudice:2012ms} 
G.~F.~Giudice, P.~Paradisi and M.~Passera,
%``Testing new physics with the electron g-2,''
JHEP {\bf 1211} (2012) 113 
[arXiv:1208.6583 [hep-ph]].

\bibitem{Aboubrahim:2014hya}
A.~Aboubrahim, T.~Ibrahim and P.~Nath,
``Probe of New Physics using Precision Measurement of the Electron Magnetic Moment,''
arXiv:1403.6448 [hep-ph].
%%CITATION = ARXIV:1403.6448;%%

\bibitem{Mohr:2012tt} 
P.~J.~Mohr, B.~N.~Taylor and D.~B.~Newell,
%``CODATA Recommended Values of the Fundamental Physical Constants: 2010,''
Rev.\ Mod.\ Phys.\  {\bf 84} (2012) 1527 
[arXiv:1203.5425 [physics.atom-ph]].

\bibitem{Passera:2004bj} 
  M.~Passera,
  %``The Standard model prediction of the muon anomalous magnetic moment,''
  J.\ Phys.\ G {\bf 31} (2005) R75 
  [hep-ph/0411168].
  %%CITATION = HEP-PH/0411168;%%

\bibitem{Aoyama:2012wk} 
  T.~Aoyama, M.~Hayakawa, T.~Kinoshita and M.~Nio,
  %``Complete Tenth-Order QED Contribution to the Muon g-2,''
  Phys.\ Rev.\ Lett.\  {\bf 109} (2012) 111808 
  [arXiv:1205.5370 [hep-ph]].
  %%CITATION = ARXIV:1205.5370;%%

\bibitem{Czarnecki:1995sz} 
  A.~Czarnecki, B.~Krause and W.~J.~Marciano,
  %``Electroweak corrections to the muon anomalous magnetic moment,''
  Phys.\ Rev.\ Lett.\  {\bf 76} (1996) 3267 
  [hep-ph/9512369].
  %%CITATION = HEP-PH/9512369;%%

\bibitem{Gribouk:2005ee} 
  T.~Gribouk and A.~Czarnecki,
  %``Electroweak interactions and the muon g-2: Bosonic two-loop effects,''
  Phys.\ Rev.\ D {\bf 72} (2005) 053016 
  [hep-ph/0509205].
  %%CITATION = HEP-PH/0509205;%%

\bibitem{Gnendiger:2013pva} 
  C.~Gnendiger, D.~Stöckinger and H.~Stöckinger-Kim,
  %``The electroweak contributions to $(g-2)_\mu$ after the Higgs boson mass measurement,''
  Phys.\ Rev.\ D {\bf 88} (2013) 5,  053005
  [arXiv:1306.5546 [hep-ph]].
  %%CITATION = ARXIV:1306.5546;%%

\bibitem{Davier:2010nc} 
  M.~Davier, A.~Hoecker, B.~Malaescu and Z.~Zhang,
  %``Reevaluation of the Hadronic Contributions to the Muon g-2 and to alpha(MZ),''
  Eur.\ Phys.\ J.\ C {\bf 71} (2011) 1515 
  [Erratum-ibid.\ C {\bf 72}  (2012) 1874]
  [arXiv:1010.4180 [hep-ph]].
  %%CITATION = ARXIV:1010.4180;%%

\bibitem{Czarnecki:2001pv} 
  A.~Czarnecki and W.~J.~Marciano,
  %``The Muon anomalous magnetic moment: A Harbinger for 'new physics',''
  Phys.\ Rev.\ D {\bf 64} (2001) 013014 
  [hep-ph/0102122].
  %%CITATION = HEP-PH/0102122;%%
  
\bibitem{Biggio:2008in} 
  C.~Biggio,
  %``The Contribution of fermionic seesaws to the anomalous magnetic moment of leptons,''
  Phys.\ Lett.\ B {\bf 668} (2008) 378 
  [arXiv:0806.2558 [hep-ph]].
  %%CITATION = ARXIV:0806.2558;%%
  
\bibitem{Ilakovac:2013wfa} 
  A.~Ilakovac, A.~Pilaftsis and L.~Popov,
  %``Lepton Dipole Moments in Supersymmetric Low-Scale Seesaw Models,''
  Phys.\ Rev.\ D {\bf 89} (2014) 015001 
  [arXiv:1308.3633 [hep-ph]].
  %%CITATION = ARXIV:1308.3633;%%

\bibitem{Abdallah:2011ew} 
  W.~Abdallah, A.~Awad, S.~Khalil and H.~Okada,
  %``Muon Anomalous Magnetic Moment and mu -> e gamma in B-L Model with Inverse Seesaw,''
  Eur.\ Phys.\ J.\ C {\bf 72}  (2012) 2108
  [arXiv:1105.1047 [hep-ph]].
  %%CITATION = ARXIV:1105.1047;%%

\bibitem{Abdallah:2003xd} 
  J.~Abdallah {\it et al.}  [DELPHI Collaboration],
  %``Study of tau-pair production in photon-photon collisions at LEP
  %and limits on the anomalous electromagnetic moments of the tau
  %lepton,'' 
  Eur.\ Phys.\ J.\ C {\bf 35} (2004) 159 [hep-ex/0406010].

\bibitem{Eidelman:2007sb} 
  S.~Eidelman and M.~Passera,
  %``Theory of the tau lepton anomalous magnetic moment,''
  Mod.\ Phys.\ Lett.\ A {\bf 22} (2007) 159 
  [hep-ph/0701260].
  %%CITATION = HEP-PH/0701260;%%

\bibitem{FernandezMartinez:2007ms}
  E.~Fernandez-Martinez, M.~B.~Gavela, J.~Lopez-Pavon and O.~Yasuda,
  %``CP-violation from non-unitary leptonic mixing,''
  Phys.\ Lett.\ B {\bf 649} (2007) 427
  [hep-ph/0703098].
  %%CITATION = HEP-PH/0703098;%%

\bibitem{Capozzi:2013csa} 
  F.~Capozzi, G.~L.~Fogli, E.~Lisi, A.~Marrone, D.~Montanino and A.~Palazzo,
  %``Status of three-neutrino oscillation parameters, circa 2013,''
  Phys.\ Rev.\ D {\bf 89} (2014) 093018 
  [arXiv:1312.2878 [hep-ph]].
  %%CITATION = ARXIV:1312.2878;%%

\bibitem{Goudzovski:2011tc} 
  E.~Goudzovski [NA48/2 and NA62 Collaborations],
  %``Kaon programme at CERN: recent results,''
  PoS EPS {\bf -HEP2011} (2011) 181 
  [arXiv:1111.2818 [hep-ex]].
  
\bibitem{Lazzeroni:2012cx} 
  C.~Lazzeroni {\it et al.}  [NA62 Collaboration],
  %``Precision Measurement of the Ratio of the Charged Kaon Leptonic Decay Rates,''
  Phys.\ Lett.\ B {\bf 719} (2013) 326 
  [arXiv:1212.4012 [hep-ex]].
  
\bibitem{Naik:2009tk} 
  P.~Naik {\it et al.}  [CLEO Collaboration],
  %``Measurement of the Pseudoscalar Decay Constant f(D(s)) Using
  %D(s)+ ---> tau+ nu, tau+ ---> rho+ anti-nu Decays,'' 
  Phys.\ Rev.\ D {\bf 80} (2009) 112004 
  [arXiv:0910.3602 [hep-ex]].
  
\bibitem{Li:2011nij} 
  H.~-B.~Li,
  %``Proceedings, 4th International Workshop on Charm Physics (Charm
  %2010) : Beijing, China, October 21-24, 2010,'' 
  Int.\ J.\ Mod.\ Phys.\ Conf.\ Ser.\  {\bf 02} (2011).
  
\bibitem{Aubert:2007xj} 
  B.~Aubert {\it et al.}  [BaBar Collaboration],
  %``A Search for $B^{+} \to \tau^{+} \nu$ with Hadronic $B$ tags,''
  Phys.\ Rev.\ D {\bf 77} (2008) 011107.
  
\bibitem{Adachi:2012mm} 
  I.~Adachi {\it et al.}  [Belle Collaboration],
  %``Evidence for $B^- \to \tau^- \bar{\nu}_\tau$  with a Hadronic
  %Tagging Method Using the Full Data Sample of Belle,'' 
  Phys.\ Rev.\ Lett.\  {\bf 110}  (2013) 131801
  [arXiv:1208.4678 [hep-ex]].

\bibitem{Cirigliano:2007xi}
  V.~Cirigliano and I.~Rosell,
  %``Two-loop effective theory analysis of pi (K) ---> e anti-nu/e [gamma] branching ratios,''
  Phys.\ Rev.\ Lett.\  {\bf 99} (2007) 231801
  [arXiv:0707.3439 [hep-ph]].
  %%CITATION = ARXIV:0707.3439;%%

\bibitem{Finkemeier:1995gi}
  M.~Finkemeier,
  %``Radiative corrections to pi(l2) and K(l2) decays,''
  Phys.\ Lett.\ B {\bf 387} (1996) 391
  [hep-ph/9505434].
  %%CITATION = HEP-PH/9505434;%%
  
\bibitem{Atre:2009rg} 
  A.~Atre, T.~Han, S.~Pascoli and B.~Zhang,
  %``The Search for Heavy Majorana Neutrinos,''
  JHEP {\bf 0905}  (2009) 030
  [arXiv:0901.3589 [hep-ph]].
  
\bibitem{Adam:2013mnn} 
  J.~Adam {\it et al.}  [MEG Collaboration],
  %``New constraint on the existence of the $\mu^+ \to e^+\gamma$ decay,''
  Phys.\ Rev.\ Lett.\  {\bf 110} (2013) 201801 
  [arXiv:1303.0754 [hep-ex]].
  
\bibitem{Auger:2012ar} 
  M.~Auger {\it et al.}  [EXO Collaboration],
  %``Search for Neutrinoless Double-Beta Decay in $^{136}$Xe with EXO-200,''
  Phys.\ Rev.\ Lett.\  {\bf 109} (2012) 032505 
  [arXiv:1205.5608 [hep-ex]].
 
\bibitem{Albert:2014awa} 
  J.~B.~Albert {\it et al.}  [EXO-200 Collaboration],
  %``Search for Majorana neutrinos with the first two years of EXO-200 data,''
  Nature {\bf 510} (2014) 229-234 
  [arXiv:1402.6956 [nucl-ex]].
  
\bibitem{Gando:2012zm} 
  A.~Gando {\it et al.}  [KamLAND-Zen Collaboration],
  %``Limit on Neutrinoless $\beta\beta$ Decay of Xe-136 from the First
  %Phase of KamLAND-Zen and Comparison with the Positive Claim in
  %Ge-76,'' 
  Phys.\ Rev.\ Lett.\  {\bf 110} (2013) 062502 
  [arXiv:1211.3863 [hep-ex]].
  
\bibitem{DeliaTosionbehalfoftheEXO:2014zza} 
  [Delia Tosi on behalf of the EXO Collaboration],
  ``The search for neutrino-less double-beta decay: summary of current experiments,''
  arXiv:1402.1170 [nucl-ex].
    
\bibitem{Gorla:2012gd} 
  P.~Gorla [CUORE Collaboration],
  %``The CUORE experiment: Status and prospects,''
  J.\ Phys.\ Conf.\ Ser.\  {\bf 375} (2012) 042013.
  
\bibitem{Artusa:2014lgv} 
  D.~R.~Artusa {\it et al.}  [CUORE Collaboration],
  ``Searching for neutrinoless double-beta decay of $^{130}$Te with CUORE,''
  arXiv:1402.6072 [physics.ins-det].
  
\bibitem{Hartnell:2012qd} 
  J.~Hartnell [SNO+ Collaboration],
  %``Neutrinoless Double Beta Decay with SNO+,''
  J.\ Phys.\ Conf.\ Ser.\  {\bf 375} (2012) 042015 
  [arXiv:1201.6169 [physics.ins-det]].
  
\bibitem{Barabash:2012gc} 
  A.~Barabash [SuperNEMO Collaboration],
  %``SuperNEMO double beta decay experiment,''
  J.\ Phys.\ Conf.\ Ser.\  {\bf 375} (2012) 042012.
  
\bibitem{Granena:2009it} 
  F.~Granena {\it et al.}  [NEXT Collaboration],
  ``NEXT, a HPGXe TPC for neutrinoless double beta decay searches,''
  arXiv:0907.4054 [hep-ex].

\bibitem{Gomez-Cadenas:2013lta} 
  J.~J.~Gomez-Cadenas {\it et al.}  [NEXT Collaboration],
  ``Present status and future perspectives of the NEXT experiment,''
  arXiv:1307.3914 [physics.ins-det].
  
\bibitem{Wilkerson:2012ga} 
  J.~F.~Wilkerson, E.~Aguayo, F.~T.~Avignone, H.~O.~Back, A.~S.~Barabash, J.~R.~Beene, M.~Bergevin and F.~E.~Bertrand {\it et al.},
  %``The MAJORANA demonstrator: A search for neutrinoless double-beta decay of germanium-76,''
  J.\ Phys.\ Conf.\ Ser.\  {\bf 375} (2012) 042010.

\bibitem{Bulbul:2014sua}
  E.~Bulbul, M.~Markevitch, A.~Foster, R.~K.~Smith, M.~Loewenstein and S.~W.~Randall,
  %``Detection of An Unidentified Emission Line in the Stacked X-ray
  %spectrum of Galaxy Clusters,'' 
  Astrophys.\ J.\  {\bf 789} (2014) 13
  [arXiv:1402.2301 [astro-ph.CO]].
  %%CITATION = ARXIV:1402.2301;%%
  
\bibitem{Boyarsky:2014jta}
  A.~Boyarsky, O.~Ruchayskiy, D.~Iakubovskyi and J.~Franse,
  ``An unidentified line in X-ray spectra of the Andromeda galaxy and Perseus galaxy cluster,''
  arXiv:1402.4119 [astro-ph.CO].
  %%CITATION = ARXIV:1402.4119;%%

\bibitem{Gelmini:2008fq}
  G.~Gelmini, E.~Osoba, S.~Palomares-Ruiz and S.~Pascoli,
  %``MeV sterile neutrinos in low reheating temperature cosmological scenarios,''
  JCAP {\bf 0810} (2008) 029
  [arXiv:0803.2735 [astro-ph]].
  %%CITATION = ARXIV:0803.2735;%%

\bibitem{Dasgupta:2013zpn}
  B.~Dasgupta and J.~Kopp,
  %``A m\'enage \`a trois of eV-scale sterile neutrinos, cosmology, and structure formation,''
  Phys.\ Rev.\ Lett.\  {\bf 112} (2014) 031803
  [arXiv:1310.6337 [hep-ph]].
  %%CITATION = ARXIV:1310.6337;%%

\bibitem{Blennow:2010th} 
  M.~Blennow, E.~Fernandez-Martinez, J.~Lopez-Pavon and J.~Menendez,
  %``Neutrinoless double beta decay in seesaw models,''
  JHEP {\bf 1007} (2010) 096 
  [arXiv:1005.3240 [hep-ph]].

\bibitem{Abada:2014vea}
  A.~Abada and M.~Lucente,
  ``Looking for the minimal inverse seesaw realisation,''
  arXiv:1401.1507 [hep-ph].
  %%CITATION = ARXIV:1401.1507;%%

 \bibitem{GonzalezGarcia:1988rw} 
  M.~C.~Gonzalez-Garcia and J.~W.~F.~Valle,
  %``Fast Decaying Neutrinos and Observable Flavor Violation in a New Class of Majoron Models,''
  Phys.\ Lett.\ B {\bf 216} (1989) 360.
  %\cite{Casas:2001sr}

\bibitem{Casas:2001sr}
  J.~A.~Casas and A.~Ibarra,
  %``Oscillating neutrinos and muon ---> e, gamma,''
  Nucl.\ Phys.\ B {\bf 618} (2001) 171
  [hep-ph/0103065].
  %%CITATION = HEP-PH/0103065;%%

\bibitem{Forero:2011pc} 
  D.~V.~Forero, S.~Morisi, M.~Tortola and J.~W.~F.~Valle,
  %``Lepton flavor violation and non-unitary lepton mixing in low-scale type-I seesaw,''
  JHEP {\bf 1109} (2011) 142 
  [arXiv:1107.6009 [hep-ph]].
  %%CITATION = ARXIV:1107.6009;%%

\bibitem{Malinsky:2009gw}
  M.~Malinsky, T.~Ohlsson and H.~Zhang,
  %``Non-unitarity effects in a realistic low-scale seesaw model,''
  Phys.\ Rev.\ D {\bf 79} (2009) 073009
  [arXiv:0903.1961 [hep-ph]].
  %%CITATION = ARXIV:0903.1961;%%

\bibitem{Dev:2009aw}
P.~S.~B.~Dev and R.~N.~Mohapatra,
%``TeV Scale Inverse Seesaw in SO(10) and Leptonic Non-Unitarity Effects,''
Phys.\ Rev.\ D {\bf 81} (2010) 013001
[arXiv:0910.3924 [hep-ph]].
%%CITATION = ARXIV:0910.3924;%%

\bibitem{Kopp:2013vaa}
  J.~Kopp, P.~A.~N.~Machado, M.~Maltoni and T.~Schwetz,
  %``Sterile Neutrino Oscillations: The Global Picture,''
  JHEP {\bf 1305} (2013) 050
  [arXiv:1303.3011 [hep-ph]].
  %%CITATION = ARXIV:1303.3011;%%
  

  \bibitem{Abada:2014zra} 
  A.~Abada, G.~Arcadi and M.~Lucente,
  ``Dark Matter in the minimal Inverse Seesaw mechanism,''
  arXiv:1406.6556 [hep-ph].
  
  
\end{thebibliography}
\end{document}